\documentclass[preprint]{emulateapj}
\usepackage{graphicx}
\usepackage{lscape}

\def\DMAG{${\Delta M_{\rm min}}$}
\def\CZ{${\Delta({\rm c} z_{\rm max}})$}
\def\DMAX{${D_{\rm max}}$}
\newcommand{\gsim}{\hbox{ \rlap{\raise 0.425ex\hbox{$>$}}\lower 0.65ex\hbox{$\sim$} }}
\newcommand{\lsim}{\hbox{ \rlap{\raise 0.425ex\hbox{$<$}}\lower 0.65ex\hbox{$\sim$} }}
\def\met{$\rm [Z/H] $}
\def\en{$\rm [\alpha/Fe] $}

\shorttitle{On the Nature of Fossil Groups}
\shortauthors{La Barbera et al.}
\journalinfo{Accepted to AJ}
\begin{document}

\title{On the Nature of Fossil Galaxy Groups: Are they really fossils ?}

\author{La Barbera, F.\altaffilmark{1}, de Carvalho, R.R.\altaffilmark{2, 3}, de la Rosa, I.G.\altaffilmark{4}, Sorrentino, G.\altaffilmark{1}, Gal, R.R.\altaffilmark{5}, Kohl-Moreira, J.L.\altaffilmark{6}} 

\altaffiltext{1}{INAF-Osservatorio Astronomico di Capodimonte, Via Moiariello 16, 80131 Napoli, ITALY}
\altaffiltext{2}{VSTceN, via Moiariello 16, 80131 Napoli, ITALY}
\altaffiltext{3}{INPE/DAS Av. dos Astronautas 1758, S\~ao Jos\'e dos Campos, SP Brazil}
\altaffiltext{4}{Instituto de Astrofisica de Canarias, Tenerife, Spain}
\altaffiltext{5}{Institute for Astronomy, 2680 Woodlawn Dr., Honolulu, HI 96822}
\altaffiltext{6}{Observat\'orio Nacional, Rua General Jos\'e Cristino 77, Rio de Janeiro Brazil}

\begin{abstract}
We  use SDSS-DR4 photometric  and spectroscopic  data out  to redshift
$z\sim0.1$ combined with ROSAT All  Sky Survey X-ray data to produce a
sample of  twenty-five fossil groups  (FGs), defined as  bound systems
dominated by a single,  luminous elliptical galaxy with extended X-ray
emission.   We  examine  possible  biases introduced  by  varying  the
parameters  used  to define  the  sample  and  the main  pitfalls  are
discussed.  The  spatial density of  FGs, estimated via  the $V/V_{\rm
  MAX}$ test, is 2.83 $\times$10$^{-6} h_{75}^3$ Mpc$^{-3}$ for L$_{X}
> 0.89 \! \times \!  10^{42} h_{75}^{-2}$ erg s$^{-1}$ consistent with
Vikhlinin  et   al.   (1999),  who  examined   an  X-ray  overluminous
elliptical galaxy sample (OLEG).  We compare the general properties of
FGs  identified  here  with  a  sample  of  bright  field  ellipticals
generated  from   the  same  dataset.   These  two   samples  show  no
differences in  the distribution  of neighboring faint  galaxy density
excess, distance from the red sequence in the color-magnitude diagram,
and  structural   parameters  such  as  a$_{4}$   and  internal  color
gradients.  Furthermore, examination of stellar populations shows that
our   twenty-five   FGs   have   similar  ages,   metallicities,   and
$\alpha$-enhancement as the  bright field ellipticals, undermining the
idea that these systems represent fossils of a physical mechanism that
occurred at  high redshift.  Our  study reveals no  difference between
FGs and field ellipticals, suggesting that FGs might not be a distinct
family of true fossils, but rather the final stage of mass assembly in
the Universe.

\end{abstract}

\section{Introduction}

Searching for  preserved remnants of physical  processes that occurred
in the  cosmic past is one  of cosmologists' main  tools in developing
our  understanding of how  galaxies, groups,  and clusters
were formed and evolved, and ultimately came to define the large scale
structure we observe. Recent studies have shown that there seems to be
a  class of  systems with  exceptional preservation,  known  as fossil
groups (FGs, Ponman  et al. 1994). These systems  consist of isolated,
luminous early-type  galaxies embedded in an extended  X-ray halo. The
main motivation for seeking such  galactic systems stems from the fact
that the merger  time scales for $\rm L > L^*$  group galaxies is much
shorter than the cooling time  scales for the hot gas component within
which these bright galaxies are  embedded (e.g. Barnes 1989, Ponman \&
Bertram 1993). In  this simplistic view FGs are  a natural consequence
of the merging  process in normal groups and therefore  can be used to
trace the history of coalescence in the not-so-high-redshift Universe.

From the observational viewpoint, FGs are seen in the optical as large
elliptical galaxies,  but with X-ray luminosities  comparable to those
of an entire group of galaxies ($L_X > 4.4 \times 10^{42} ergs/s$; see
Jones et  al.  2003,  hereafter JO03).  To  date, no clear  picture of
their origin has emerged from  the collected data.  Two main scenarios
have been suggested: FGs result  from the complete merging of galaxies
that constituted  a loose group  in the past,  collapsing at an
early epoch, but never becoming incorporated into clusters (Hausman \&
Ostriker 1978;  Ponman et al.  1994; Jones et al.   2000; Khosroshahi,
Jones \& Ponman  2004); or they originate from  a region that inhibits
formation of  $L^*$ galaxies  in these groups  leading to  an atypical
galaxy luminosity function (LF, Mulchaey \& Zabludoff 1999).

These  systems  have  been  detected  out to  redshifts  of  at  least
$z\sim0.6$ (Ulmer et al.~2005),  have high $M/L$ ratios suggesting low
star formation efficiency (Vikhlinin et al. 1999) and represent 8\% to
20\%  of all  systems with  similar X-ray  luminosities.  Ten  FGs are
already known and  well studied from optical and  X-ray data (Mulchaey
\& Zabludoff 1999, Romer et al. 2000, hereafter R00, Jones et al 2003,
Khosroshahi,  Jones, \&  Ponman~2004, Sun  et al.   2004,  Yoshioka et
al. 2004, Ulmer et al.  2005).  A recent paper by Santos et al.~(2007)
presents a  sample of 34  FG candidates based  on SDSS data.   In this
work, we  use optical (SDSS) and  X-ray (ROSAT All Sky  Survey -- RASS)
data to  define a sample of  FG candidates. We select  FGs following a
strategy similar  to Jones et  al.  (2003), using  spectroscopic (age,
metallicity, $\rm \alpha/Fe$  enhancement) and photometric ($\rm a_4$,
color  gradient) parameters  to  further constrain  what  should be  a
FG. The main  goal is to establish a link  between FGs, compact groups
and isolated  ellipticals based on their global  properties and simple
expectations,  if  mergers are  the  dominant  events determining  the
evolution of a FG.

This  paper is  organized as  follows. Section  2 presents  the galaxy
catalog  used to  search for  FGs, describing  the parameters  and the
X-ray measurements that define the search. The actual selection of FGs
is  then detailed.  Section  3  presents the  control  sample used  to
compare  FG properties  to  those of  "normal"  ellipticals, which  in
principle  did  not  result  from  merging. Section  4  discusses  the
magnitude distribution of the  brightest and second brightest galaxies
of FGs,  while in  Section 5 we  characterize the distribution  of the
faint galaxies around  FGs.  Since FGs are thought  to be old systems,
in Section 6 we examine the colors of FGs relative to the red sequence
of early type  galaxies.  Section 7 deals with  the internal structure
of  seed galaxies  of  FGs,  as  measured by  their  structural
parameters  and internal  color gradients.  In Section  8,  we analyze
stellar population properties of  FGs, by looking at the distributions
of age, metallicity, and  $\alpha$ enhancement.  Finally, in Section 9,
we summarize  and discuss the  main results.  Throughout the  paper we
use a cosmology  with $\Omega_{\rm m} = 0.3$,  $\Omega_\Lambda = 0.7$,
and $\rm H_0 = 75 \, \rm km \, s^{-1} \, Mpc^{-1}$.

%

\section{Selection of FGs}
FGs are defined as bound  systems of galaxies associated with extended
X-ray  sources.  The  total  optical luminosity  of  these systems  is
dominated by a  bright elliptical galaxy, and their  LF exhibits a gap
between the  first and  second rank galaxies  (see e.g.~\citealt{JONES
  03}).  The sample  of FG candidates defined here  is selected on the
basis of  spectroscopic and photometric  data from the Data  Release 4
(DR4) of the Sloan Digital Sky Survey (SDSS), and on the basis of RASS
X-ray    imaging.   The    optical   selection    is    described   in
Sec.~\ref{subsec:SEL}, while the X-ray measurements and the definition
of the FG sample are presented in Sec.~\ref{subsec:XRAY}.

\subsection{The  SDSS catalog and the optical selection parameters}
\label{subsec:SEL}

The  optical  selection  is  performed on  a  complete  volume-limited
catalog   of    91563   galaxies,   retrieved    from   the   SDSS-DR4
database~\footnote{\footnotesize  http://www.sdss.org/DR4} through the
CasJobs                                facility~\footnote{\footnotesize
  http://casjobs.sdss.org/casjobs/}.   The  catalog  consists  of  all
galaxies with  absolute $r$-band magnitudes $M_r$  brighter than $-20$
and  with spectroscopic  redshifts  between  0.05 and  0.095.
Absolute  magnitudes are  obtained  from the  SDSS $r$-band  Petrosian
magnitudes, corrected for galactic extinction using the reddening maps
of~\citet{SFD98}.  The  lower redshift limit of the  catalog is chosen
to minimize the  aperture bias~\citep{GOMEZ03}, which strongly affects
large  nearby galaxies, while  the upper  redshift limit  guarantees a
high  level of completeness,  estimated through  Schmidt's $V/V_{max}$
test (see~\citealt{SAR06}). The magnitude limit of $-20$ corresponds approximately 
to the apparent magnitude limit of the SDSS spectroscopy  at 
redshift $z \sim 0.095$ ($r \sim 17.8$).

Since FGs are systems characterized by a gap in the galaxy LF, we
first select optical candidates by searching the SDSS spectroscopic
database for pairs of galaxies whose magnitude difference is larger
than a given value,~\DMAG. For each galaxy in the catalog, we
select all objects whose projected distance on the sky from the target
galaxy is smaller than a maximum radius, ~\DMAX, and whose redshift
difference is smaller than a given value, ~\CZ. In other words, for
each galaxy, we select companion galaxies in a cylinder centered on
the galaxy, with a radius of ~\DMAX \, and a semi-height of ~\CZ \,
along the line of sight.  The value of~\DMAX \, is fixed in physical
units (Mpc), and the selection is done by transforming the value
of~\DMAX \, to the apparent size that corresponds to the redshift of
the seed galaxy. A bright galaxy is defined as a possible FG if (i)
there is at least one companion selected through the above procedure,
and (ii) if all the companions are fainter, with the magnitude
difference greater than \DMAG.  Thus, our optical FG candidates are
selected on the basis of three parameters: ~\CZ, ~\DMAX, and ~\DMAG.
The parameters ~\CZ \, and ~\DMAX \, are used to select the companions
of a given galaxy in the SDSS catalog, while the parameter ~\DMAG \,
characterizes the gap in the LF between the galaxy and its companions.

\begin{figure}
\begin{center}
\plotone{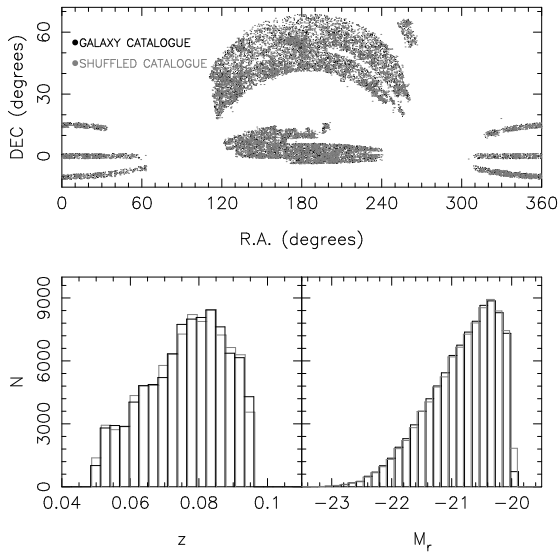}
\caption{Comparison of the characteristics  of the galaxy catalog with
  those  of one  of the  shuffled catalogs.   The upper  panel plots
  declination versus right ascension and compares the sky area covered
  from both  catalogs.  The  lower left and right panels  show the
  redshift  and magnitude distributions,  respectively, of  galaxies in
  both catalogs.  As shown in the upper panel, galaxies from the
  shuffled and  the galaxy  catalogs are plotted  with grey  and black
  colors,  respectively.}
\label{fig:CONF_DR4_SHUFFLE}
\end{center}
\end{figure}

 To determine suitable values of~\CZ,~\DMAX  \, and~\DMAG, we
examine how the number of  optical FG candidates and the corresponding
contamination  rate change  as a  function of  these  parameters.  The
parameters are chosen in such a  way that the contamination rate is at
most  15\%, defined as  the fraction  of FG  candidates that  would be
selected (for a given set  of values of ~\CZ,~\DMAX \, and~\DMAG) from
a random  distribution of galaxies  with the same mean  galaxy surface
density, magnitude, and redshift distributions as the SDSS catalog.  We
generated randomized galaxy catalogs by applying the shuffling method,
as  described in~\citet{GAL03}.   First, we  bin the  SDSS  catalog in
$4^{\circ}$ in right ascension and declination and $\Delta z=0.005$ in
redshift.   The  randomized  catalog  is then  obtained  by  shuffling
galaxies  in each bin  by randomly  exchanging their  right ascension,
declination and  redshift.  The above  bin sizes are chosen  such that
both  the  mean  surface  density  and the  redshift  distribution  of
galaxies in  the SDSS catalog  are preserved, without  overly reducing
the number  of galaxies in each  bin.  We verified  that varying these
bin sizes by $20\%$ does  not change the contamination rate estimates.
Fig.~\ref{fig:CONF_DR4_SHUFFLE}  compares the  characteristics  of the
galaxy  catalog with  those  of  one of  the  shuffled catalogs.   The
shuffled  catalogs reproduce  well the  sky coverage  of  the SDSS-DR4
catalog  as   well  as  the  corresponding   distributions  of  galaxy
magnitudes and  redshifts.  Since  the shuffling procedure  washes out
small scale structures from the galaxy catalog, the FG candidates that
are selected from the shuffled  catalogs only arise as rare and random
instances in the magnitude, redshift, and spatial distributions of the
galaxies,  instead of a population  of objects that follow  from some
physical   evolutionary  path.       To   reduce  statistical
fluctuations  in  the  contamination  rate estimates,  we  repeat  the
shuffling procedure  100 times and estimate the  contamination rate as
the ratio of the number of optical FG candidates to the mean number of
candidates  selected  from the  shuffled  catalogs  applying the  same
~\CZ,~\DMAX, and~\DMAG \, criteria.

\begin{figure}
\begin{center}
\plotone{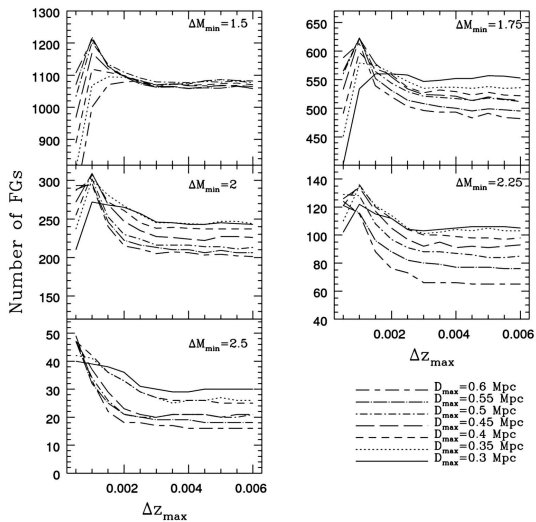}
\caption{Number  of  FG candidates  as  a  function  of the  selection
  parameter ~$\Delta z_{\rm max}$.  The panels correspond to different
  values  of ~\DMAG (shown  in the  upper right  of each  plot), while
  different line  styles correspond to different values  of ~\DMAX, as
  shown in the lower right corner of the figure.}
\label{fig:NUM_FGS_PAR}
\end{center}
\end{figure}

\begin{figure}
\begin{center}
\plotone{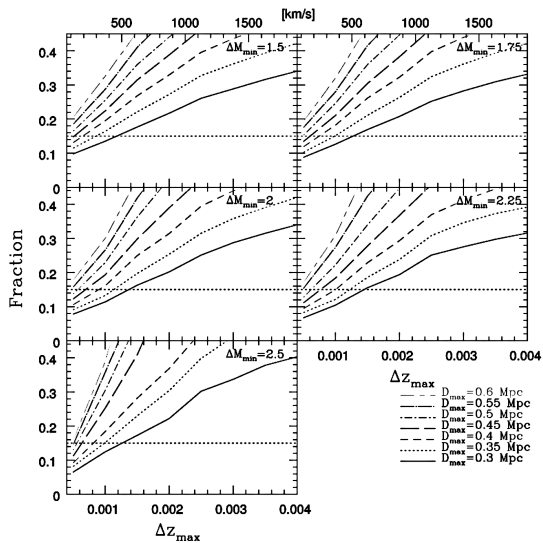}
\caption{Same as Fig.~\ref{fig:NUM_FGS_PAR} but showing the
  corresponding contamination rates as estimated through the shuffling
  method (see the text).  In each panel, the dotted horizontal line
  marks the upper limit for the acceptable level of contamination in
  the sample of FG candidates.}
\label{fig:CONT_FGS_PAR}
\end{center}
\end{figure}

\begin{figure}
\begin{center}
\plotone{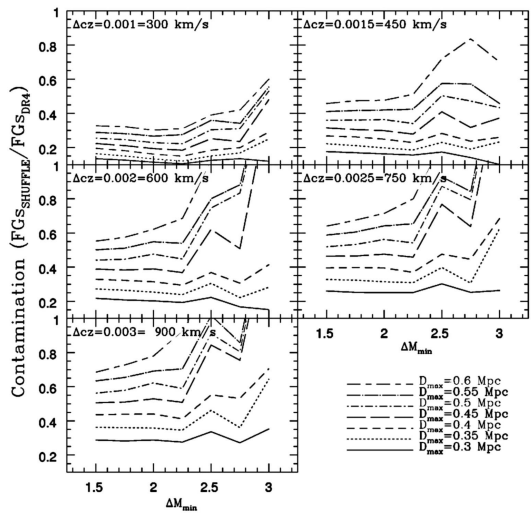}
\caption{Contamination  rate  as  a  function of  ~\DMAG.   Each  plot
  corresponds to a different value of~\CZ, as shown in the upper-right
  of  each  panel.   Curves  with  different line  styles
  correspond to  different values  of ~\DMAX, and  are the same  as in
  Fig.~\ref{fig:NUM_FGS_PAR}.   }
\label{fig:CONT_DMAG}
\end{center}
\end{figure}

\begin{figure}
\begin{center}
\epsscale{0.8}
\plotone{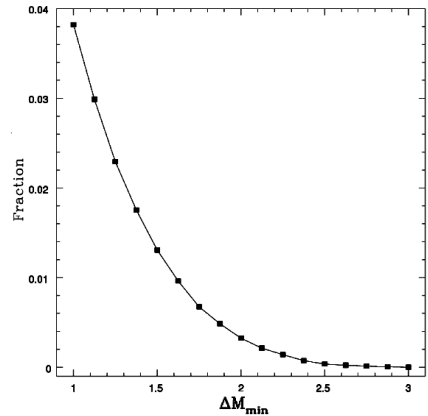}
\caption{Fraction of FG  candidates in the SDSS catalog  as a function
  of  ~\DMAG  ~for  the  values of  ~\DMAX$=0.35$Mpc  and  ~\CZ$=0.001$
  adopted to select FG candidates.  The black curve is obtained from a
  cubic interpolation of the data.}
\label{fig:NUM_DMAG}
\end{center}
\end{figure}

Figs.~\ref{fig:NUM_FGS_PAR} and~\ref{fig:CONT_FGS_PAR} show the number
of  FG optical  candidates  and the  contamination  rate as  functions
of~\CZ,~\DMAX \, and~\DMAG.  We vary ~\CZ \, from $0.00025$ ($\sim 75$
km  s$^{-1}$),  corresponding  to  $\sim  \! 2.5$  times  the  typical
redshift  uncertainty of  the SDSS  spectroscopic sample~\footnote{For
  the SDSS-DR4 main galaxy sample, the typical redshift accuracy is 30
  km   s$^{-1}$    (see   http://www.sdss.org/dr4/)},   to   $0.0065$,
corresponding to $\sim \! 2000$ km s$^{-1}$, i.e.,  about two times the
typical  velocity  dispersion of  a  rich  cluster  of galaxies.   The
minimum value  of~\DMAX \,  is chosen to  be $0.3$Mpc, which  is about
twice  the  typical  core  radius  of  galaxy  clusters~\citep{AMK98}.
Furthermore,  we  consider  only   values  of~\DMAX  \,  smaller  than
$0.6$Mpc, since  higher values yield contamination  rates greater than
$\sim  20 \%$ for  all possible  values of  the other  two parameters.
Finally, we  consider only values  of~\DMAG ~ in  the range of  1.5 to
2.5~mag.  The value of 1.5 is  chosen to have a reasonable minimum gap
in the  LF of  FG candidates,  while values of  ~\DMAG \,  larger than
$2.5$ are excluded since they  overly reduce the number of candidates.
Looking  at  Fig.~\ref{fig:NUM_FGS_PAR}, we  see  that  the number  of
candidates  shows a maximum  at \CZ$\sim  0.001$ for  a wide  range of
values of  both ~\DMAX \, and  ~\DMAG.  Hence, to  maximize the sample
size, we adopt this value of~\CZ.  For \DMAX, we see that for fixed
values of \CZ \, and \DMAG \, the number of candidates does not change
significantly for  $ 0.35 \lesssim$  \DMAX $\lesssim 0.5$Mpc.   On the
other hand, the  contamination rate changes by a factor  of 2 over the
full range of  values of \DMAX.  We choose a  suitable value of ~\DMAX
\,  by fixing  an upper  limit to  the contamination  rate.  Examining
Fig.~\ref{fig:CONT_FGS_PAR} and considering all the values of \DMAG \
that  correspond to  ~\CZ$  = 0.001$,  contamination  rates less  than
15$\%$  are  achieved only  for  \DMAX$\lesssim  0.35$Mpc.  Since  the
number of  FG candidates  decreases as a  function of \DMAX,  we adopt
\DMAX$=0.35$Mpc.  Finally,  we have to determine the  choice of \DMAG.
Fig.~\ref{fig:CONT_DMAG} plots the contamination rate as a function of
\DMAG \, for all possible values of  \DMAX \, and \CZ. We see that the
contamination  is  almost  constant  for \DMAG$\lesssim2.2$,  with  an
increase  for larger  values of~\DMAG.   Fig.~\ref{fig:NUM_DMAG} plots
the number of FG candidates as  a function of \DMAG \, for \CZ$=0.001$
and ~\DMAX$=0.35$Mpc.   The number of FG  candidates rapidly decreases
as  ~\DMAG \,  increases, reaching  almost 0  at \DMAG$\sim  2.5$.  In
order to  obtain a statistically  significant number of  FG candidates
without  overly  reducing  the  \DMAG  \, gap,  we  decided  to  adopt
\DMAG$=1.75$~mag.

To summarize, the values of the three parameters ~\CZ, ~\DMAX \, and
~\DMAG \, are chosen as a compromise between the number of selected FG
candidates and the corresponding contamination rates.  We adopt the
following parameters: \CZ$=0.001$, ~\DMAX$=0.35$Mpc \, and
~\DMAG~$=1.75$mag, resulting in a list of 578 optical FG
candidates. Note that the ~\DMAX \, value is close
to half of the virial radius (0.37 $\rm h_{75}^{-1} Mpc$) assuming a
temperature of $1 \, Kev$ and the mean redshift of the
sample~\citep{EMN96}.  As found by \citet{Khosroshahi:07}, based on
Chandra data, a typical temperature for FGs is $1$~Kev, with a few
systems having a temperature as high as $\sim 3$~Kev. In the latter
case, half the virial radius is $0.64$~Mpc. Using such a
large ~\DMAX \, results in a high contamination rate for whatever
value of~\CZ \, and ~\DMAG \, (see Fig.~\ref{fig:CONT_DMAG}).

We note that the  above  selection  of FG  systems  can  be  affected by  the  SDSS
spectroscopic incompleteness,  because of  the SDSS fiber  collision limit
preventing  neighboring fibers  from being  closer than  55$''$.  This
prevents complete spectroscopic coverage  of objects in dense regions,
even in cases  where multiple tiles overlap. It  is therefore, possible
to have  galaxies adjacent to  some optical FG candidates  which would
nominally disqualify them, but lacking SDSS spectroscopy, the
neighbors would not be taken  into account by our selection procedure.
This issue is addressed in Sec.~\ref{sec:CONTAMINATION}.

\subsection{X-ray Analysis and  the Final Sample of FGs}
\label{subsec:XRAY}
\subsubsection{Measuring X-ray Fluxes with RASS}

The X-ray luminosity L$_{X}$ of each optical FG candidate is estimated
using                                                              RASS
countrates\footnote{ftp://ftp.xray.mpe.mpg.de/rosat/archive/900000}. The
first  step is to  estimate the  background contribution,  by randomly
selecting  one  hundred  boxes  each   with  an  area  of  100  square
arcminutes,  located   throughout  the  $6.5^{\circ}\times6.5^{\circ}$
field containing  the target,  and then taking  the median and  rms of
these  hundred  values.   The  count  rates are  integrated  in  three
apertures, with  radii of  5, 10, and  20 times the  effective radius,
r$_{\rm e}$,  of the FG  seed galaxies.  Effective radii  are measured
from     the    $r$-band    SDSS     images,    as     described    in
App.~\ref{subsec:STRUCPAR}. We  considered a detection  as significant
when the S/N is at least 3-$\sigma$ above the background, otherwise we
derive  upper limits  as  in  Beuing et  al.~(1999).  We selected  for
further analysis only those systems with a significant X-ray detection
in at least one of the three measured apertures, resulting in a sample
of 113 (out of 578) FG candidates.

To convert the measured total count rate into an unabsorbed X-ray flux
in  the ROSAT 0.5  - 2.0  keV energy  band, we  use the  HEASARC PIMMS
tool. We assume  a Raymond-Smith spectrum to represent  the hot plasma
present in the intracluster medium  (Raymond \& Smith 1977) with solar
metallicity  and an  interstellar  hydrogen column  density along  the
line-of-sight given by Bajaja et al.~(2005). The plasma temperature is
fixed at 1  keV, which is typical for  very bright elliptical galaxies
(Beuing  et al.~1999),  although the  dependence  of the  flux on  the
temperature used  is insignificant  ($<5\%$). Finally, the  rest frame
0.5-2.0  keV flux  was obtained  by  applying the  K-correction as  in
B\"ohringer et al.~(2000).

\subsubsection{Contamination by  Spirals, AGN, and Clusters}
\label{sec:CONTAMINATION}
To refine the selection of FGs, we considered four possible sources of
contamination:  (i) spiral  galaxies; (ii)  AGN;  (iii) superpositions
with rich  clusters of galaxies.

(i) Since FGs are dominated by a bright elliptical galaxy, we removed 
contaminating spiral galaxies from the list of optical candidates.  We 
fit the SDSS $r$-band image of each galaxy with a seeing convolved 
Sersic model (see App.~\ref{subsec:STRUCPAR} for details), and flagged 
as spirals those objects with a detectable spiral arm pattern in the 
residual map obtained after model subtraction.  This procedure eliminates 
ninety-one (out of 578) contaminating systems.

(ii) We flagged those FG optical candidates with AGN spectral signatures and excluded them from the analysis.  AGN were selected following the criteria described in~\citet{SAR06}. We removed fourty-five (out of 578) galaxies.

The  above  selections  leave  a  sample of  102  FG  candidates  with
significant  X-ray  detections,  no  significant  AGN  signatures,  and
early-type morphology.


(iii) The resolution and S/N  of RASS is insufficient to differentiate
between X-  ray emission from a real  FG and that from  a rich cluster
which happens to be along the line of sight.  Moreover, for
a rich  cluster, the SDSS  spectroscopic incompleteness at  the bright
end might mimic the lack of bright companion galaxies required to be a
FG.  Thus,  we  rejected FG  candidates  within  1.5  Mpc (at  the  FG
redshift)  from a  rich  Abell  cluster.  We  discarded  $R >0$  Abell
clusters with the exception  of Abell~690 (richness $R=1$; FG\,$22$ of
Tab.~2)  for  which  we  verified  that there  are  no  galaxies  with
available  photometry  from SDSS  that  might  invalidate our  adopted
fossilness definition.   Twenty-eight (out of 102)  FG candidates were
discarded  with this criterion,  leaving a  sample of  seventy-four FG
candidates.

\begin{figure}[]
\begin{center}
\epsscale{0.5}
\plotone{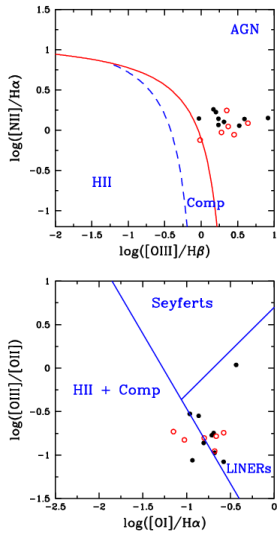}
\caption{Diagnostic  diagrams showing the  nine FG  candidates (filled
  circles)  which   still  show  some  emission  lines   in  the  SDSS
  spectra. The empty circles represent the six field galaxies (FS; see
  Sec.~\ref{sec:FS_FS_CONF}) with  some residual emission  features in
  the SDSS  spectra.  We also  display the regions defining  the locus
  for AGN (Seyfert  and LINERs) and starbursts (HII).   Notice that the
  figure also shows the transition region between AGN and HII (Comp).
 \label{fig:BPT}}
\end{center}
\end{figure}

Although  we eliminated  AGN and  spiral galaxies  from the  sample, we
further  investigated the presence  of elusive  emission lines  in the
SDSS spectra  of the 102  FGs with significant X-ray  detections.  For
each spectrum, we  modeled the absorption lines with  a combination of
stellar population  models using the STARLIGHT code  (Cid Fernandes et
al.  2005).  Further  details   of  the  procedure  are  presented  in
Sec.~\ref{sec:SP}.  After subtracting  the old  stellar  component, we
measured   the   remaining   $O_{\rm   I}$,  $O_{\rm   II}$,   $O_{\rm
  III}$,$H_{\rm \alpha}$,  $H_{\rm \beta}$, and  $N_{\rm II}$ emission
features.  Fig.~\ref{fig:BPT} displays the diagnostic diagrams defined
with these indices. Regions corresponding to different types of active
galaxies     are    plotted     according     to    the     definition
of~\citet{Kew06}. Only nine FG candidates have residual emission lines
typical of AGN. In  agreement with expectations from our morphological
selection,  no   starburst  galaxies  (HII)   are  found  in   the  FG
sample.  Moreover, almost  all of  the AGN  are classified  as LINERs,
which   have  a   typical   X-ray  luminosity   of  $10^{40}$   ergs/s
(see~\citealt{Kom99}, two orders of  magnitudes lower than the minimum
X-ray luminosity of the FGs in  our sample. Thus, we conclude that AGN
and starburst  contamination does not contribute  significantly to the
X-ray  measurements, except  for AGN  that might  be detected  only in
X-rays. However,  as noted by~\citet{AND07}, these account  for only a
minor fraction of all AGN.


\subsubsection{How extended is the X-ray emission?}

An  important  issue  to   consider  is  whether  the  observed  X-ray
luminosity comes  from an extended source.  Although  AGN were excised
from our sample following the recipes in~\citet{SAR06}, some AGN might
only be  detected in X-ray (\citealt{TOZZI06,  AND07}), representing a
further  source  of  contamination  in  the FG  sample.    To
objectively  establish the extended  nature of  the X-ray  emission we
proceed as follows:

\begin{figure}[]
\begin{center}
\epsscale{1.0}
\plotone{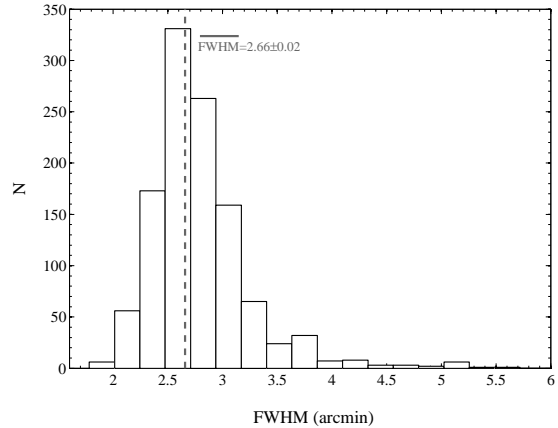}
\caption{ Distribution  of FWHM values  for 1232 point sources  in the
  RASS  Bright Source  Catalog.  The vertical  dashed  line marks  the
  position of the peak of  the distribution. The location of the peak,
  $\overline{FWHM}$, and the corresponding $1\sigma$ error are shown in
  the upper part of the figure. \label{fig:stars}}
\end{center}
\end{figure}

1 - We  select 1232 point sources from the  RASS Bright Source Catalog
(\citealt{Voges99}),   requiring  that   the  sources   have   a  high
probability  of  being  a   real  detection  (i.e.   source  detection
likelihood $>100$) and a source  extension of zero.  This data set was
used to measure the mean FWHM of  the RASS PSF over the whole sky.  We
used 2DPHOT  (La Barbera et al  2008) to fit a  Moffat distribution to
every  source, after  a gaussian  smoothing  (with a  $\sigma$ of  one
pixel)  of the  original count  rate  image.  The  smoothing was  done
similarly  for the  FGs (see  below).  Fig.~\ref{fig:stars}  shows the
distribution of the  FWHM for the 1232 point  sources. The location of
the peak  of the  distribution (and its  error) was obtained  from the
biweight   estimator   (\citealt{Beers:90})  and   is   found  to   be
$2.66$($\pm0.02$)~arcmin. The narrowness  of the distribution confirms
the  expectations that  the  RASS PSF  has  little off-axis  variation
resulting from  the scanning strategy.  We also  verified that fitting
the  unsmoothed images  gives a  mean FWHM  of $2.15$~arcmin,  in good
agreement with the value of $2.1$~arcmin from~\citet{degrandi97}.

2 - For the seventy-four FGs, we smoothed the original RASS images
with a 1-pixel Gaussian (as for the point sources). Source detection
was performed using S-Extractor with a detection threshold of
2$\sigma$ over an area of 5 pixels, using a tophat detection filter,
which is optimal for faint source detection.  The gaussian smoothing
enhances the contrast over the background. For each FG, we identified
the closest X-ray source to the SDSS position.

3 - Using the smoothed X-ray images, we fit a Moffat distribution to
the FG candidates.  Here, the error in the FWHM was estimated from
one-hundred Monte Carlo simulations. For each simulation, all the
detected sources were masked from the original RASS image and the
remaining pixels were bootstrapped to create a random background image,
which was then gaussian smoothed. The Moffat model was added to this
image, and the fitting was repeated. We subtracted in quadrature the
mean FWHM of the RASS smoothed PSF from the FWHM of each source. This
quantity was defined as the source extension, with its error computed
from the uncertainty on the FWHM.

4 - We matched X-ray and optical sources by considering only cases
where the distance between the optical and X-ray FG positions were
smaller than the FWHM of the X-Ray source.  Only
fourty-three (out of seventy-four) FG candidates have X-ray
counterparts meeting this requirement.

\begin{figure}[]
\begin{center}
\epsscale{1.0}
\plotone{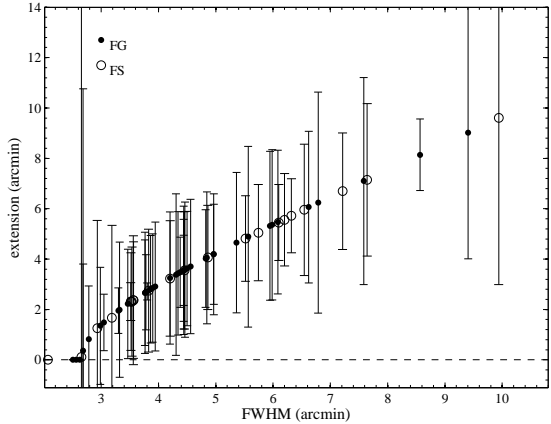}
\caption{ Extension parameter versus FWHM for the 43 FG candidates
  (filled circles) with an associated X-ray source. Empty circles
  represent the sample of twenty-two field galaxies defined in
  Sec.~\ref{sec:FS_FS_CONF}.  The error bars show 2$\sigma$ standard
  uncertainties. Sources are defined as extended if their error bar
  does not cross the horizontal dashed line corresponding to a
  zero extension.
\label{fig:ext}}
\end{center}
\end{figure}

5 - We classified the FG X-ray sources as extended if the extension
parameter was greater than 0 at the 2$\sigma$ level (see
Fig.~\ref{fig:ext}). This selection removes eight point-like FG
candidates, leading to a sample of thirty-five extended FGs. In the
following section, we discuss the effect of SDSS spectroscopic
incompleteness on this sample.

\subsubsection{Spectroscopic incompleteness and the final sample}
\label{sec:spec_incomp}

The  SDSS spectroscopy is  incomplete in  high-density regions  due to
limitations  on  fiber   placement.   We  address  this  spectroscopic
incompleteness  issue  as follows:  for  each  of  the thirty-five  FG
candidates, we select all galaxies with SDSS photometry fulfilling the
\DMAX \, criterion and with  magnitude between $m$ and $m+1.75$, where
$m$ is  the FG $r$-band  Petrosian magnitude.  These are  the galaxies
that could  possibly invalidate  the fossilness definition.   Then, we
queried  the SDSS  DR6 and  NED for  spectroscopic redshifts  of these
possible contaminants. We found that  for nine (out of thirty-five) FG
candidates,  there is  at least  one  galaxy invalidating  the \CZ  \,
criterium,   hence,  disqualifying   the  system.    Using  photometric
redshifts,   we  also   rejected  one   system  with   three  possible
contaminants having  concordant redshifts within  2$\sigma$.  Applying
the same analysis to the  eight point-like FG candidates (see previous
section) leaves four such objects having no possible gap contaminants.

\begin{figure}[]
\begin{center}
\epsscale{0.9}
\plotone{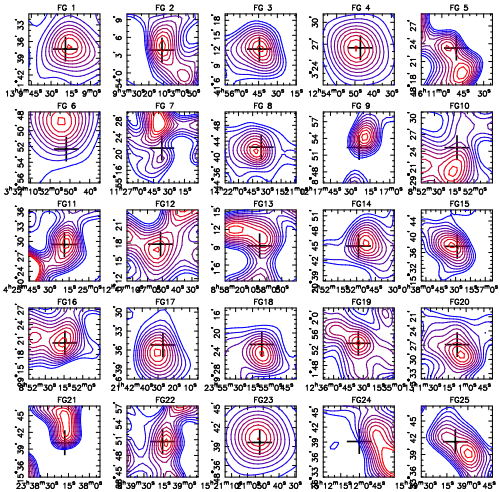}
\caption{X-ray  emission contours  from the  RASS for  the twenty-five
  FGs, with the position of the  optical SDSS galaxy image marked by a
  black cross.  The contours have  been obtained by applying a 2 pixel
  gaussian  smoothing to  the original  RASS images  and  are equally
  spaced at  intervals of 1$\sigma$ starting at  0.5$\sigma$ above the
  background.  The reddest  contours mark  the centroid  of  the X-ray
  emission. Each panel is three times as wide as the FWHM of the X-ray
  source.
 \label{XRAYMAPS}}
\end{center}
\end{figure}

After  this  procedure,  our  final  sample  consists  of  twenty-five
systems\footnote{We  note that  almost all  FGs in  our sample  have a
  total  X-ray luminosity  above the  (bolometric) limit  of  $\rm 4.4
  \times 10^{42} \,  ergs/s$ adopted by Jones et  al.(2003).  Only two
  FGs  (FG\,11 and  FG\,21, see  Tab.~2) have  an X-ray  luminosity of
  $\sim \rm 2 \times 10^{42} \, ergs/s$, corresponding to a bolometric
  value of $\sim \rm 3.5 \times  10^{42} \, ergs/s$.}, which we use to
characterize the properties of FGs.  This is the largest sample of FGs
available    in    the   low    redshift    regime   ($z<0.1$).     In
Fig.~\ref{XRAYMAPS},  we show the  X-ray contour  plots for  these FGs
with the optical SDSS position overlaid. As we can see, usually,
the  centroid  of  the  X-ray  component coincides  with  the  optical
counterpart.
\subsection{Archive data and external comparison}

We  searched the  XMM, Chandra,  ASCA  and Einstein  archives for  any
publicly available data,  taken with any of the  instruments, within a
1.5$h_{75}^{-1}$  Mpc radius  of the  final samples  of FGs  and field
galaxies (defined in Sec.~\ref{sec:FS_FS_CONF}).  We found data in the
vicinity of  only a  single FG  in the Chandra  archive.  None  of the
other  archives contained  any observations  including objects  in our
final sample. We conclude that there are no useful archival data deeper
than the RASS to systematically examine the X-ray properties of our FG
and field galaxies, or exclude them from the sample as X-ray AGN.

We also looked for FGs defined in the literature that are in the same
redshift interval as our sample. Not surprisingly, we found only three
systems well studied in several contributions
(e.g.~JO03,~\citealt{Khosroshahi:07}), although none of them have data
in the SDSS. Moreover, \citet{Santos07}, defining a sample of
thirty-four FGs from SDSS-DR4, find only four systems in the same
redshift regime as we consider here. One of them is in our sample of
FG candidates, but with an X-ray component indistinguishable from the
RASS PSF. The other three systems are not in our FG sample because the
absolute magnitude limit of our SDSS catalog prevents finding a
second-rank galaxy for the main targets.

\subsection{The Space Density of FGs}

The space density of the twenty-five FGs was estimated using the $1 /
V_{\rm max}$ statistics suggested by Avni \& Bahcall (1980).  The sky
area is that of the DR4, 4783 square degrees.  Applying various X-ray
luminosity limits we find different integrated space densities, which
we compared to those of previous studies.  Vikhlinin et al. (1999,
hereafter V99) studying an overluminous elliptical galaxy (OLEG)
sample found a space density of 1.5 $\times$10$^{-6} h_{75}^3$
Mpc$^{-3}$ for L$_X > 0.44 \times 10^{42} h_{75}^{-2}$ erg/s while we
find 3.4 $\times$10$^{-6} h_{75}^3$ Mpc$^{-3}$. For the same limiting
value of L$_{X}$, Romer et al.~(2000) find 6.7 $\times$10$^{-6}
h_{75}^3$ Mpc$^{-3}$, JO03 finds 6.75 $\times$10$^{-7} h_{75}^3$
Mpc$^{-3}$, and \citet{dariush:07} predicts 5.4 $\times$10$^{-6}
h_{75}^3$ Mpc$^{-3}$ based on the Millennium simulation.  Considering
the errors in these measurements, as presented in Table~1 of JO03, we
can conclude that all these values are consistent.  When a brighter
limiting L$_{X}$ is considered, L$_{X} > 0.89 \times 10^{42}
h_{75}^{-2}$ erg/s, V99 finds 8.1 $\times$10$^{-6} h_{75}^3$
Mpc$^{-3}$ and we find 2.83 $\times$10$^{-6} h_{75}^3$ Mpc$^{-3}$,
still consistent.  We emphasize that the samples analyzed by R00, V99,
and JO03 have only 3, 4, and 5 systems, respectively, while we have
twenty-five new FGs, yet the space densities are still very
consistent.

\section{Comparing the properties of FGs and Field Galaxies}
\label{sec:FS_FS_CONF}
\begin{figure}
\begin{center}
\epsscale{1.0}
\plotone{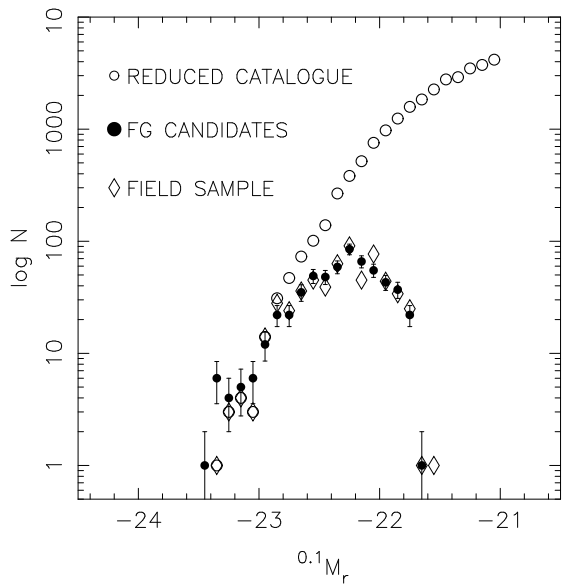}
\caption{LFs  of  galaxies in  the  reduced  catalog (empty  circles),
  brightest galaxies in FG candidates (filled circles) and of galaxies
  in the field sample (diamonds). The symbols used for each sample are
  shown in  the upper  left part  of the figure.  The error  bars mark
  1$\sigma$ Poissonian uncertainties on  number counts of FG brightest
  galaxies.}
\label{fig:LF_RS}
\end{center}
\end{figure}

To establish  a benchmark  for the properties  of FGs, we  generated a
control sample  of galaxies with the same  luminosity distribution and
the same optical  and X-ray properties as the  FG sample.  The control
sample represents  an idealization of  a sample of systems  which were
probably  formed by a  single collapse  of a  protogalaxy and  did not
experienced any major mergers in  last 3--4 Gyrs.  From now on, we refer
to the control sample as 'field' galaxies. This was created by
first excluding  the brightest galaxies  of the FG  optical candidates
and their companions (see Sec.~\ref{subsec:SEL}) from the whole galaxy
catalog, yielding a reduced catalog with 90259 targets.  We binned the
sample of 578 FG optical  candidates and the reduced galaxy catalog by
$r$-band  absolute magnitude,  and  for each  bin  computed the  ratio
between the number  $N_1$ of FG optical candidates in  that bin to the
number  $N_2$ of  galaxies in  the reduced  catalog.   Galaxy absolute
magnitudes  were computed  as described  in App.~\ref{subsec:RS_MEAS}.
 To obtain a sample of  'field' galaxies with the  same LF as
that  of  the seed  galaxies  of the  578  optical  FG candidates,  we
randomly extracted galaxies from the reduced catalog in each magnitude
bin  according to the  corresponding value  of the  ratio $f=N_1/N_2$.
Because we  have excluded the  brightest galaxies and  their companion
galaxies in the optical FG  candidates from the entire galaxy catalog,
it  is  possible  for  $N_2$  to  be lower  than  $N_1$,  yielding  an
ill-defined  value of  $f$.   This is  shown in  Fig.~\ref{fig:LF_RS},
where we  compare the LF of  the brightest galaxies in  the optical FG
candidates  with  that  of   galaxies  in  the  reduced  catalog.   At
magnitudes brighter than  $M_{\rm r} \sim -23$, the  number of optical
FG candidates in each magnitude bin is slightly larger than the number
of galaxies in  the reduced catalog.  To ensure that  the value of $f$
is  always  less than  one,  we enlarged  the  size  of the  brightest
magnitude bins such that the condition $N_2<N_1$ was always fulfilled.
This goal was  achieved by considering the same  magnitudes bins as in
Fig.~\ref{fig:LF_RS} for $M_{\rm r}  \gsim -22.8$, and just one
bin at  brighter magnitudes,  from $M_{\rm r}  = -23.6$ to  $M_{\rm r}
\sim -22.8$.   The list of  field galaxies was extracted  according to
this binning procedure,  and, by construction, it has  the same sample
size  ($N=578$) as the  optical FG  candidates.  Fig.~\ref{fig:LF_RS},
which  also plots  the LF  of  this field  sample, shows  that its  LF
actually  matches  the LF  of  the seed  galaxies  in  the optical  FG
candidates.  As  expected from the fact  that we use  a single, larger
magnitude  bin at  $M_{\rm r}  \lesssim -23$,  there is  only  a small
difference between the two LFs in this magnitude range.  However, this
difference is only marginally significant, provided that uncertainties
on number counts are taken into account.

 To  perform  a  detailed comparison  between  FGs and  field
galaxies, we  extracted a subsample  of field galaxies,  following the
same procedure  adopted to select the  sample of FGs  from the initial
list of  578 optical  FG candidates (see  Sec.~\ref{subsec:XRAY}).  We
removed  spirals and  AGN  from  the catalog  of  field galaxies,  and
analyzed the X-ray images of  the remaining systems. Sixty-six (out of
578) field  galaxies have  significant X-ray detections.   From these,
following the prescriptions outlined  in the previous section, we kept
only the twenty-two galaxies associated  with an X-ray source and with
a   projected  distance   from   rich  Abell   clusters  larger   than
1.5Mpc. Applying the X-ray extent requirement leaves a final sample of
seventeen  field  galaxies, shown  in  Fig.~\ref{fig:ext}.  A  similar
analysis  to   that  performed  in  Sec.~2.2.2  for   FGs  shows  that
contamination from  elusive AGN and starbursts is  also negligible for
the sample  of field galaxies  (see Fig.~6). Although selected  in the
same  way from an  optical catalog  with the  same number  of galaxies
(578) and the same galaxy LF as the FGs, the number of galaxies in the
field sample with significant X-ray detections ($N=66$) is $65\%$ less
than the  corresponding FG sample  ($N=102$).  The final FG  and field
samples have a similar ratio of $68\%$.

In  the following sections,  we compare  the distributions  of several
properties   (luminosity,   structural   parameters,  internal   color
gradients, stellar  populations) of FG  and field galaxies.   For each
quantity,  the comparison  is  performed (i)  by a  Kolmogorov-Smirnov
(hereafter KS)  test and (ii) by  deriving the mean and  width of each
distribution.  The  mean and width  values obtained for  each quantity
and the results of the KS tests are reported in Tab.~\ref{tab:MWCONF}.
In order to  reduce the effects of outliers  in each distribution, the
corresponding mean, and width  values were computed using a $2.5\sigma$
clipping  method.  We also  verified that  computing the  location and
width  values with ROSTAT  (Beers et  al.  1990)  does not  affect our
results. All quantities  used in this analysis are  reported in Tab.~2
for each  FG and field galaxy  (see App.~A). Although not  used in the
present analysis, the table also  lists the properties of the four FGs
with point-like  X-ray emission that  survived the final  section step
described   in  Sec.~\ref{sec:spec_incomp}.    The   distributions  of
photometric and X-ray properties  of FGs and field galaxies (hereafter
also   referred  as   the   field  sample,   FS)   are  presented   in
Fig.~\ref{fig:MAG_FG_FS} and described in the following sections.

\begin{deluxetable*}{c|cc|cc|c}
\tiny
\tablecaption{Mean and width values of the distributions of several quantities for FG and FS
  galaxies. The p-values from 
  the  KS  test comparing the  distributions of  FG and  FS  properties
  are reported in Col.(6). Values smaller than 0.1  indicate that the two samples likely  do   not    derive   from   the   same    parent   distribution.
}
\tabletypesize{\tiny}
\tablecolumns{5}
\tablehead{ & \multicolumn{2}{c}{$FG$} &  \multicolumn{2}{c}{$FS$} & KS \\
 & \colhead{mean} & \colhead{width} & \colhead{mean} & \colhead{width}  &  \\
 (1) & (2) & (3) & (4) & (5) & (6) } 
\startdata
  $^0.1M_r$                                                         & $-22.64\pm 0.09$ & $ 0.37\pm 0.07$ & $-22.45\pm 0.09$ & $ 0.37\pm 0.05$ & $  0.36 $ \\
  $\delta_{\rm N}$                                                  & $  2.5\pm  0.4$ & $ 1.56\pm 0.30$ & $  2.5\pm  0.5$ & $  2.0\pm  0.4$ & $  0.99 $ \\
  $\log R_{\rm e} \, \rm (kpc)$                             & $ 1.13\pm 0.08$ & $ 0.29\pm 0.06$ & $ 0.97\pm 0.08$ & $ 0.35\pm 0.10$ & $  0.23 $ \\
  $n$                                                               & $  5.4\pm  0.6$ & $  2.4\pm  0.6$ & $  5.5\pm  0.6$ & $  2.5\pm  0.4$ & $  0.83 $ \\
  $a_{4}*100$                                                       & $ 0.11\pm 0.26$ & $  0.6\pm  0.3$ & $  0.3\pm  0.2$ & $ 0.75\pm 0.11$ & $  1.00 $ \\
  $\nabla(g-r)$                                             & $-0.072\pm0.026$ & $ 0.09\pm 0.06$ & $-0.17\pm 0.08$ & $ 0.25\pm 0.08$ & $  0.74 $ \\
$\log L_{\rm X} (\cdot 10^{44} h_{75}^{-2} \, \rm ergs/s)$    & $-0.90\pm 0.09$ & $ 0.41\pm 0.13$ & $-0.84\pm 0.15$ & $ 0.52\pm 0.11$ & $  0.44 $ \\
  $\delta_{\rm CM}$                                               & $  0.0\pm  0.3$ & $  0.8\pm  0.4$ & $ -0.5\pm  0.3$ & $  1.1\pm  0.2$ & $  0.29 $ \\
  $\log \sigma_0 \, \, (km/s)$                                                    & $ 2.43\pm 0.01$ & $ 0.07\pm 0.01$ & $ 2.42\pm 0.04$ & $ 0.09\pm 0.05$ & $  0.40 $ \\
  $\log Age \, \, (Gyr)$                                                  & $ 0.72\pm 0.03$ & $ 0.14\pm 0.03$ & $ 0.77\pm 0.04$ & $ 0.16\pm 0.04$ & $  0.10 $ \\
  $[Z/H]$                                                   & $ 0.44\pm 0.03$ & $ 0.17\pm 0.02$ & $ 0.34\pm 0.06$ & $ 0.24\pm 0.05$ & $  0.29 $ \\
  $[\alpha/Fe]$                                                       & $ 0.27\pm 0.02$ & $ 0.10\pm 0.01$ & $ 0.28\pm 0.03$ & $ 0.12\pm 0.02$ & $  0.93 $ \\
\enddata
\label{tab:MWCONF}
\end{deluxetable*}

\section{Luminosities of first and second rank galaxies}
\label{sec:MAG_DIST}
In this section, we analyze the properties of FGs by
examining  the magnitude  distributions of  their seed and second
rank galaxies as well as their X-ray fluxes.

\begin{figure}
\begin{center}
\epsscale{1.0}
\plotone{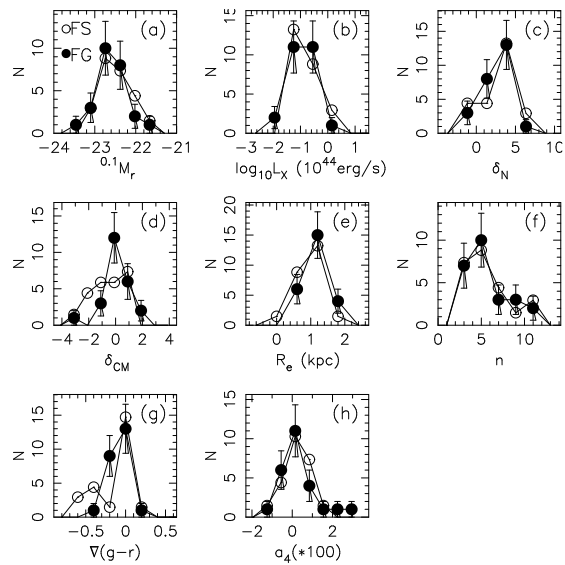}
\caption{Distributions  of  FG  and  FS  properties  as  follows:  (a)
  absolute  magnitude, (b)  X-ray luminosity,(c)  density  excess, (d)
  normalized  distance  to  the  red sequence,  (e)  r-band  effective
  radius, (f)  r-band Sersic index,  (g) internal color  gradient, (h)
  boxiness parameter. FG and field  galaxies are plotted as filled and
  empty  circles, respectively  (see upper--left  panel).   Error bars
  mark  the $1  \sigma$  standard confidence  intervals  of FG  number
  counts.   To make  the plot  more readable,  uncertainties  on field
  galaxy  number counts  are  not shown.   The  distribution of  field
  galaxies  is normalized  to  the total  number  of corresponding  FG
  galaxies.
\label{fig:MAG_FG_FS}}
\end{center}
\end{figure}

Fig.~\ref{fig:MAG_FG_FS}a    compares     the    optical    luminosity
distributions   of  FG  first   rank  galaxies   with  those   of  the
corresponding  field  sample.   The  mean  and width  values  of  each
distribution are  reported in Tab.~\ref{tab:MWCONF}.   Both the figure
and the table show that the  $r$-band LFs of FG and field galaxies are
indistinguishable.  This result is also  confirmed by the KS test (see
Tab.~\ref{tab:MWCONF}),     whose    p-value    is     $\gsim    0.3$.
Fig.~\ref{fig:MAG_FG_FS}b shows that not only the optical but also the
X-ray luminosities of  FG galaxies are consistent with  those of field
galaxies.   This is  also  clear from  the  mean, width,  and KS  test
results reported  in Tab.\ref{tab:MWCONF}. The fact that  FG and field
galaxies have  the same X-ray luminosities is  of particular interest,
since it implies that both samples consist of galaxies embedded within
halos of  the same  mass. Hence, any  difference between FG  and field
samples is not related to the  selection of these samples, but is more
deeply  related either  to the  initial  conditions of  the galaxy  LF
within the halos, or to the different evolutionary history of galaxies
within the halo.

\begin{figure}
\begin{center}
\epsscale{1.0}
\plotone{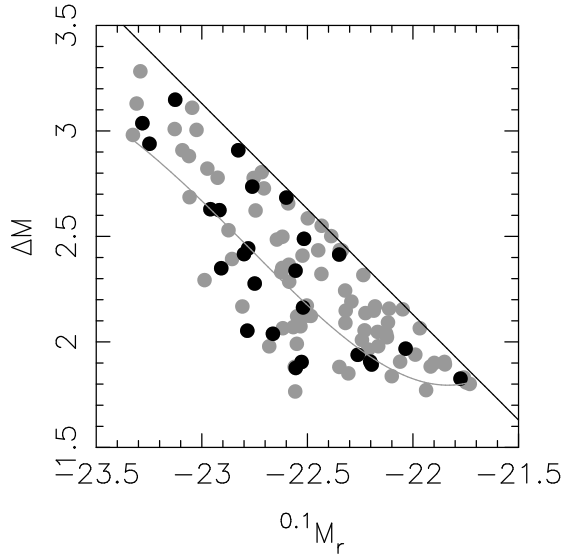}
\caption{ Differences between $r$-band  magnitudes of second and first
  rank galaxies of  FGs, $\Delta M$, are plotted as  a function of the
  absolute magnitude,  $^{0.1}M_{\rm r}$, of the  first rank galaxies.
  Grey  circles denote  the sample  of  102 FG  candidates with  X-ray
  significant  detection  (see  Sec.~\ref{subsec:XRAY}),  while  black
  symbols  correspond to  the final  sample of  twenty-five  FGs.  The
  black line  marks the  range of  allowed values of  $\Delta M$  as a
  function of $^{0.1}M_{\rm  r}$, while the grey curve  shows the mean
  $\Delta M$  value as  a function of  $^{0.1}M_{\rm r}$,  obtained by
  randomly extracting second rank  galaxies from the SDSS catalog (see
  the text).  }
\label{fig:DMAG_MAG}
\end{center}
\end{figure}

  To  analyze  the  luminosity distribution  of  second  rank
galaxies,  we have  to consider  that the  SDSS-DR4 galaxy  catalog is
magnitude  limited and  that  FG optical  candidates  are selected  as
systems  where  the magnitude  difference  between  the two  brightest
galaxies is greater than ~\DMAG.   This implies that a larger fraction
of  FGs can  be selected  at the  brightest magnitudes  of  the galaxy
LF. In other  words, the seed galaxies of FGs do  not form a magnitude
complete    sample.    Such    incompleteness   is    illustrated   in
Fig.~\ref{fig:DMAG_MAG},  where  we   plot  the  magnitude  difference
$\Delta M$  between the  second and  first rank galaxies  of FGs  as a
function of the $r$-band absolute magnitude, $^{0.1}M_{\rm r}$, of the
first rank  galaxy.  Absolute magnitudes are obtained  as described in
App.~\ref{subsec:RS_MEAS},  by  k-correcting  the  Petrosian  $r$-band
magnitudes of  the DR4 dataset to  redshift $z \!  =  \!  0.1$.  Since
the SDSS galaxy  catalog is limited to $M_{\rm  r}=-20$, and since, by
definition, we have $\Delta M > $\DMAG$=1.75$mag, the seed galaxies of
FGs have to fulfill the equation
\begin{equation}
^{0.1}M_{\rm  r} +^{0.1}k_r< -20-  \Delta M \sim -21.75 {\rm mag}.   
\label{EQM}
\end{equation}
The term $^{0.1}k_r$ denotes the k-correction computed as described in
App.~\ref{subsec:RS_MEAS}, and is  introduced in the previous equation
to account  for the  fact that k-corrections  were not applied  in the
computation  of the $r$-band  absolute magnitudes  used to  select the
SDSS-DR4  catalog.  Since the  typical value  of $k_r$  is $-0.13$mag,
Eq.~\ref{EQM}  implies   the  constraint  $^{0.1}M_{\rm   r}  \lesssim
-21.62$, and in  fact, as shown in Fig.~\ref{fig:DMAG_MAG},  we do not
find  any  first  rank  galaxy  fainter  than  this  magnitude  limit.
Moreover, for a  given value of $^{0.1}M_{\rm r}$,  we can only select
FGs  with  $^{0.1}M_{\rm r}  +^{0.1}k_r +  \Delta M  < -20$,
which  implies $  \Delta M  < -^{0.1}M_{\rm  r} -20  -  ^{0.1}k_r \sim
-^{0.1}M_{\rm r} -19.87 $.  This  last constraint is shown by the line
in Fig.~\ref{fig:DMAG_MAG},  and results in a larger  range of allowed
values of $\Delta  M$ \, at bright galaxy  magnitudes. In other words,
the selection procedure affects  the magnitude distribution of FGs, 
biasing the sample towards a larger allowed fraction of
systems  at   brighter   magnitudes.   Despite   this   bias,
Fig.~\ref{fig:DMAG_MAG} shows  that the brightest  first rank galaxies
do not show a larger range  of $\Delta M$ values, although they are
allowed. At bright magnitudes, we see that seed galaxies of FGs 
are characterized by a larger magnitude gap. However, one has to note 
that the SDSS spectroscopic incompleteness due to fiber collisions  might
cause an overestimate of $\Delta M$, and that  the SDSS spectroscopic survey 
did not target many very bright galaxies ($r<15.5$), which might reduce
the number of detected FG systems at the bright end.

 To  better understand  the distribution  of $\Delta  M$ with
respect  to the  magnitude of  the seed  galaxies, we  ran Monte-Carlo
simulations.   For each  seed  galaxy  of the  102  FG candidates,  we
assigned  a fictitious  second rank  galaxy from  the SDSS  catalog by
randomly extracting a galaxy with  $\Delta M \ge $~\DMAG.  We repeated
this procedure several times ($N=10000$), and estimated the mean value
of   $\Delta   M$   obtained    for   the   randomized   second   rank
galaxies.  Fig.~\ref{fig:DMAG_MAG}  shows that  this  mean $\Delta  M$
value actually decreases as $^{0.1}M_{\rm r}$ increases. At magnitudes
fainter  than $^{0.1}M_{\rm  r} \sim  -23$, the  simulated  $\Delta M$
value is  fully consistent  with the $\Delta  M$ distribution  of FGs,
implying that the $\Delta M$ vs.  $^{0.1}M_{\rm r}$ correlation can be
simply  explained by  the shape  of the  LF of  second  rank galaxies.
However, at bright magnitudes ($^{0.1}M_{\rm  r} < -23$), all the seed
galaxies of  FGs have $\Delta  M$ values larger than  the simulations,
which  might reflect some  intrinsic physical  property of  FG systems
(see Sec.~\ref{sec:DISCUSSION}).


\section{Faintest Galaxies: density excess}
\label{sec:DENS_EX}
We characterize the population of the faintest galaxies around each FG
by measuring the density excess  of these galaxies with respect to the
field population.  Since  galaxies around FGs can be  fainter than the
magnitude  limit of  the SDSS  spectroscopic catalog,  we  measure the
density excess  by taking advantage of the  SDSS-DR4 photometric data,
whose   completeness   limit\footnote{This   is  defined   as   95$\%$
  detectability   for  point-like   sources.}   is  $\sim22.2$mag   in
$r$-band~\citep{SLB02}.   The  density  excess, $\delta_{\rm  N}$,  is
defined as follows:
\begin{eqnarray}
\delta_{\rm N} &=&\frac{  \rho_{\rm in} - \rho_{\rm out}  }
        {\sqrt{\sigma_{\rm in}^2+\sigma_{\rm out}^2}},
\end{eqnarray}
where $\rho_{\rm in}$ is the mean galaxy density within a inner circle
of radius $r_{\rm  in}$ centered on the FG  seed galaxy and $\rho_{\rm
  out}$ is the mean galaxy  density in a concentric annulus with inner
and outer radii of  $r_{1}$ and $r_{2}$, respectively.  The quantities
$\sigma_{\rm  in}$   and  $\sigma_{\rm  out}$   denote  the  1$\sigma$
uncertainties  on  $\rho_{\rm in}$  and  $\rho_{\rm  out}$, which  are
estimated by accounting  for Poisson noise in the  galaxy counts.  For
each FG, we measure $\delta_{\rm N}$ by setting $r_{\rm in}=0.25R_{\rm
  Abell}$,  $r_{\rm  1}=1R_{\rm  Abell}$ and  $r_{\rm  2}=2R_{Abell}$,
respectively, where $R_{Abell}$ is  the apparent size corresponding to
one Abell radius ($1.5$Mpc) at the redshift of the FG seed galaxy.  We
have  verified that our  results do  not change  when these  radii are
varied by as much  as 20\%.  To probe a fixed range  of the galaxy LF,
we measure $\rho_{\rm in}$ and $\rho_{\rm out}$ using only galaxies in
the magnitude range of $m_2$  to $m_2+2$, where $m_2$ is the magnitude
of the second rank galaxy in each FG.


Fig.~\ref{fig:MAG_FG_FS}c  compares the $\delta_{\rm  N}$ distributions
of FGs  and field galaxies.   Since we do
not have second  rank galaxies for the field  sample, we obtained the
corresponding density excesses as follows.  For each field galaxy with
apparent and absolute magnitudes $m_{\rm r}$ and $^{0.1}M_{\rm r}$, we
randomly   assign  a   magnitude  shift   $\Delta  M$   following  the
distribution of  magnitude differences  between first and  second rank
galaxies of the 102 FG candidates   (see
Fig.~\ref{fig:DMAG_MAG}).  Then, we  set $m_2=m_{\rm r}+\Delta M$, and
measure  the density  excess  around the  field  galaxy following  the
procedure outlined above. In other  words, we assign a fictitious {\it
  second rank} galaxy to each  field galaxy according to the magnitude
distribution  of first  and  second rank  galaxies  in the sample of 102 FG  candidates.
Fig.~\ref{fig:MAG_FG_FS}c shows that most  galaxies in each sample tend
to  have  positive $\delta_{\rm  N}$  values.  In  fact, as  shown  in
Tab.~\ref{tab:MWCONF}, the  mean value of the density  excess is about
$2.5$ for  both samples, and, considering  the corresponding uncertainty,
is always significantly larger than  zero.  This implies that there is
an excess of faint galaxies in  the surrounding regions of both FG and
field galaxies.   Both Fig.~\ref{fig:MAG_FG_FS}c  and the  mean and  width values  reported in
Tab.~\ref{tab:MWCONF} show that  the distributions of $\delta_{\rm N}$
values of FG  and field galaxies are fully consistent. 

\section{FGs and the Red Sequence}
\label{subsec:DIST_RS}
The  distribution of  FG seed  galaxies in  color-magnitude  space can
provide clues  to their formation  history. We characterize  the $g-r$
vs.  $r$ color-magnitude relation  of early-type galaxies as described
in App.~\ref{subsec:RS_MEAS}, and then  compare the distribution of FG
and field  galaxies to  that relation. It  is implicit in  this simple
analysis   that   the   red   sequence  mainly   originates   from   a
mass--metallicity   relation  (see  e.g.~\citealt{KoA98}),   and  that
galaxies lying  on the sequence  represent, as a  first approximation,
passively evolving  systems that have not  experienced recent episodes
of star  formation.  By  measuring how distant  the colors of  FGs are
from the color-magnitude  relation, we have an indication  of how star
formation is evolving in these systems.

We compute the distance $\delta_{CM}$ of the colors of FG and field
galaxies to the red sequence as
\begin{equation}
\delta_{CM} = \frac{\left[ (g-r) - (a+ b \times \, ^{0.1}M_r) \right] }{\sigma_{g-r}(^{0.1}M_r)},
\label{eq:DIST_RS}
\end{equation}
where $a$  and $b$  are the offset  and slope of  the color--magnitude
relation, and  $\sigma_{g-r}(^{0.1}M_r)$ is the scatter  about the red
sequence   (see   Eq.~\ref{SGR}   of   Appendix~\ref{subsec:RS_MEAS}).
Fig.~\ref{fig:MAG_FG_FS}d  compares the $\delta_{CM}$  distribution of
FG and  field samples,  with the corresponding  mean and  width values
tabulated  in  Tab.~\ref{tab:MWCONF}.  The  results  of  the KS  test,
comparing  the $\delta_{CM}$  distributions, is  also reported  in the
same table.   We see that  the $\delta_{CM}$ distributions  are always
peaked near zero, as one would expect since FGs and field galaxies are
morphologically            selected            as           early-type
systems. Fig.~\ref{fig:MAG_FG_FS}d suggests that field galaxies have a
broader distribution than FGs.  However, the corresponding mean values
are fully  consistent with zero  and their widths are  consistent with
unity,  implying no significant  difference between  the $\delta_{CM}$
distributions.  This result is also  confirmed by the p-value from the
KS test.


\section{Structural properties and color gradients}
We analyze the surface brightness  distribution of FG seed galaxies by
studying (i) their structural parameters and internal color gradients,
and  (ii) the departures  of galaxy  isophotes from  purely elliptical
shapes.  Internal  color gradients are estimated  using the structural
parameters  of galaxies,  specifically the  effective  radius, $r_{\rm
  e}$, and the  Sersic index (shape parameter), $\rm  n$, while galaxy
isophotes are  characterized using the $a_4$  parameter that describes
boxy/disky    departures   of    the    isophotes   from    elliptical
shapes~\citep{Bender:87}.   The  derivation  of  these  quantities  is
described in App.~\ref{subsec:STRUCPAR}.

Figs.~\ref{fig:MAG_FG_FS}e    and~\ref{fig:MAG_FG_FS}f   compare   the
distributions of effective radii and shape parameters for FG and field
galaxies. The mean  and width values of FGs and  the field sample (see
Tab.~\ref{tab:MWCONF}) are fully consistent,  with the p-values of the
KS tests greater than $\sim 0.2$.
The lack  of galaxies with Sersic  index $n \gsim  2$ is a
natural consequence of  our sample consisting of morphologically selected
early-type  galaxies. Although  the Sersic  index does  not  provide a
precise measurement  of the bulge  to disk ratio  of a galaxy,  it can
effectively separate  disk and bulge dominated  systems, with $n\sim2$
as     the    dividing     value    between     the     two    classes
(e.g.~\citealt{Blanton:05}).

Fig.~\ref{fig:MAG_FG_FS}g   compares   the   distributions  of   color
gradients, $\nabla(g-r)$,  for FG and  field galaxies. 
We  find that fossils and  field galaxies  have a consistent  mean color  gradient of
$\nabla(g-r)  \sim -0.07$, but  the scatter  of field  galaxies around
the mean value  is somewhat larger than that  of FGs (see Tab.~\ref{tab:MWCONF}). The  
difference between the width  values of the  two samples is $0.16 \pm 0.1$,  which is 
(marginally) greater than zero by $1.6 \sigma$. The KS test (Tab.~\ref{tab:MWCONF}) confirms that 
there is no significant difference between the two samples.  Our  color  gradient
estimates can be directly compared with those of~\citet{LaBarbera:05},
who derived the  mean  $g-r$ color gradient in early-type galaxies in
clusters of different richnesses at different redshifts.  For galaxies
in poor  groups at a  redshift of $z  \sim 0.08$, close to  the median
redshift of our FG and FS samples, they found a mean color gradient of
$-0.064^{+0.008}_{-0.02}$  (see their Tab.~1),   agreeing with the
mean $\nabla(g-r)$ values  reported in Tab.~\ref{tab:MWCONF}. The mean
value  of $\nabla(g-r)$  has  also  been measured  from  SDSS data  by
~\citet{Wu:05},  who analyzed a  sample of  36 early-type  galaxies at
$z\sim 0.02$. They find a mean $\nabla(g-r)$ value of $-0.05 \pm 0.01$
(see  their Tab.~3),  which differs  by  less than 2$\sigma$  from the
results  in Tab.~\ref{tab:MWCONF}.   

In Fig.~\ref{fig:MAG_FG_FS}h, we compare the distributions of $a_4$
values for FG and field galaxies. Looking at the mean and width values
reported in Tab.~\ref{tab:MWCONF} as well as at the KS test results
reported in Tab.~\ref{tab:MWCONF}, the distributions for FG, and field
galaxies are again fully consistent. The distribution of $a_4$ values
carries interesting clues about the formation of FG galaxies.
Analyzing the isophotal shapes of seven FG galaxies drawn from the
sample of Jones et al.~(2003), Khosroshahi et al.~(2006, hereafter
KPJ06) found that all FG galaxies tend to have disky isophotes. In
more detail, examining Fig.3, three (of seven) FGs in the KPJ06 sample
have $a_4$ values very close to zero, with the remaining four galaxies
having significantly larger values.  In our sample, we find that the
distribution of FGs is peaked around $a_4=0$, with equal numbers of
disky and boxy systems.  The disagreement we find with KPJ06 may
reflect the small sample analyzed in that study.  Our comparison of
FG and field galaxies shows no difference in isophotal deviations from
elliptical shapes.  In other words, FG galaxies do not have peculiar
isophotes when compared to `normal' field galaxies.

\section{Stellar Populations}
\label{sec:SP}
To  analyze the  stellar populations  of FGs  and field galaxies,  we use
spectra from the  SDSS-DR4 with a resolution of  4.42 \AA\ (FWHM). For
each galaxy, we  search for the Single Stellar  Population (SSP) model
that  best matches  the galaxy  spectrum.  The  matching  is performed
using measurements of spectral indices, resulting in estimates of age,
metallicity,  and  elemental abundance  ratios  \en  .  The  procedure
consists of  (i) estimating the spectral indices, and (ii)
extracting the stellar population parameters, as detailed in App.~\ref{app:spectra}.

\begin{figure}
\begin{center}
\epsscale{0.8}
\plotone{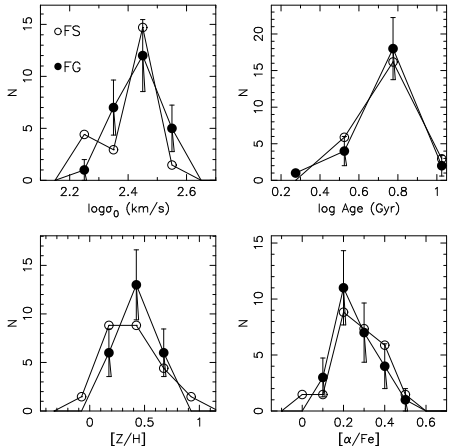}
\caption{  Same  as  Fig.~\ref{fig:MAG_FG_FS} but  comparing  velocity
  dispersions  and  stellar population  parameters  for  FG and  field
  galaxies.~\label{fig:SP_PAR} }
\end{center}
\end{figure}

\begin{figure}
\begin{center}
\epsscale{0.8}
\plotone{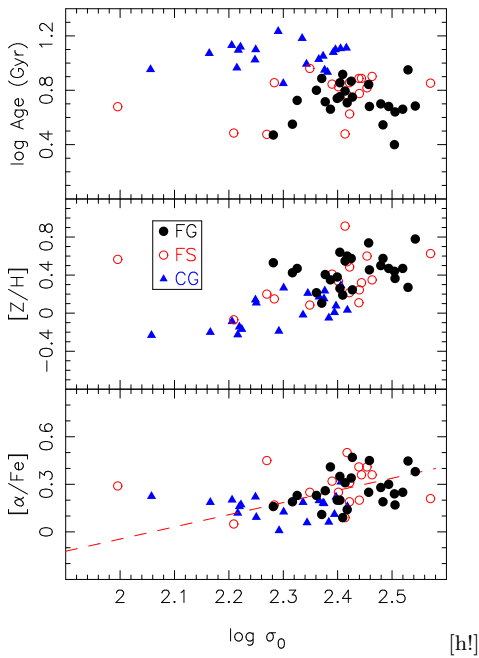}[h!]
\caption{Correlations of  stellar population parameters  with velocity
  dispersion for FGs, field  galaxies, and galaxies in Hickson Compact
  Groups, represented  with different symbols as shown  from the inset
  in  the  middle panel.  In  the  bottom figure,  as  an  aid to  the
  discussion, the best  fit to the sample of  sixty-six field galaxies
  (see Sec.~\ref{sec:FS_FS_CONF}) is plotted (dashed line).  Values of
  stellar population  parameters from  the de la  Rosa et  al.  (2007)
  have been  transformed to the $\alpha$-models system  used here (see
  App.~\ref{app:spectra}).~\label{fig:SIG_SP}}.
\end{center}
\end{figure}

Velocity  dispersions and  stellar population  parameters for  FGs and
field galaxies are reported in Tab.~2.  The distributions of $\sigma$,
$\rm age$, \met,  and \en \, are shown  in Fig.~\ref{fig:SP_PAR}, with
the  corresponding  mean and  width  values,  and  the KS  test  being
reported in  Tab.~\ref{tab:MWCONF}. We  see that the  distributions of
FGs and field samples are fully consistent, with the p-value of the KS
test  being larger  than $\sim0.1$.   Fig.~\ref{fig:SIG_SP}  shows how
stellar  population   parameters  vary  as  a   function  of  velocity
dispersion for both FGs and  field galaxies.  We see that both samples
follow the same correlation between \en \, and velocity dispersion, as
found from  previous studies for field  galaxies (e.g.  de  la Rosa et
al.~2007).   Moreover, both  FGs and  field galaxies  occupy  the same
region  in  the  diagrams  of  age  and  metallicity  versus  velocity
dispersion,   emphasizing  the   similarity  of   both   samples.   In
Fig.~\ref{fig:SIG_SP},   we  also   insert   the  stellar   population
parameters  for  the twenty  elliptical  galaxies  in Hickson  compact
groups  (HCGs) analyzed  by de  la  Rosa et  al.~(2007).  The  stellar
population parameters of these  galaxies were remeasured following the
same procedure as  for FGs and field galaxies.   Note that ellipticals
in HCGs  have higher  ages and lower  metallicities than both  FGs and
field  galaxies, showing no  correlation between  \en \,  and velocity
dispersion.

\section{Discussion}
\label{sec:DISCUSSION}

In the current hierarchical model of galaxy formation, the basic
idea is that massive galaxies form  later as the result of the merging
history  of their  dark matter  halos,  but their  stars form  earlier
(downsizing).   In this sense  special attention  is being  focused on
massive  (luminous) galactic systems,  whether they  reside in  low or
high  density locations.   We  may find  fossils  at different  cosmic
epochs,  and they  provide important  clues that  help  us distinguish
between nature and nurture.  Our main results from this paper are:

We found  578 candidates following our  optical photometric definition
of FGs in  the redshift range 0.05 $<$ z  $<$ 0.095.  After discarding
incorrect morphological  classifications and AGN we are  left with 102
FGs for which there is a significant X-ray detection in ROSAT (maximum
total flux of 10$^{-14} \rm erg \, cm^{-2} s^{-1}$).  Only twenty-five
FGs  were  selected  with  likely  extended  X-ray  emission  and  not
associated with  rich clusters. These  constitute our final  sample of
FGs. Applying  the V/V$_{\rm max}$ test  we find a  spatial density of
2.83  $\times$10$^{-6}  \rm  {h_{75}}^{3}$ Mpc$^{-3}$,  comparable  to
three other  independent observational  studies (V99, RO00,  and JO03)
and to the recent analysis  of the Millennium Simulation by Dariush et
al. (2007).

As shown  in Figure 12,  the brighter FGs  always have $\Delta  \rm M$
greater  than $\sim  3.0$ mag  while fainter  FGs are  limited  by the
completeness of  the SDSS  catalog. The lack  of systems in  the lower
left part of the figure,  at magnitudes brighter than $^{0.1}M_{\rm r}
\sim -23$,  might be  caused by  selection effects as  well as  by the
shape  of the  galaxy LF.   Thus,  it is  not clear  from the  present
analysis if the absence of systems in the lower left part of Fig.12 is
related to  some physical characteristics  of the sample. It  would be
reasonable to expect  that as mergers progress the  LF of such systems
will  have  a larger  gap  between the  first  and  the second  ranked
galaxies.

The density excess around FGs is fully consistent with what we measure
for field galaxies, implying that the environments around both systems
are  statistically   identical  when  considering   the  faint  galaxy
population. This result seems to  indicate that FGs and field galaxies
originate  from  similarly  overdense   regions  of  the  Large  Scale
Structure.

By studying the  distribution of color offsets from  the red sequence,
we find  that FGs  occupy a similar  region to field  ellipticals.  We
also find  that the mean color  gradients in both samples  (FG and FS)
are statistically the same.  However,  the scatter around the mean for
FGs  is  significantly (1.6  $\sigma$)  smaller  than  that for  field
ellipticals, which may indicate a  more regular process in the buildup
of  FGs, such  as mergers  of L$^{*}$  galaxies.  The  distribution of
$a_{4}$,  which  measures  isophotal  shape  deviations  from  a  pure
ellipses, indicates that FGs and field ellipticals are similar.

Studying the stellar populations  in the twenty-five FGs and seventeen
field galaxies we find that there is no difference in age, metallicity
or $\alpha$-enhancement, indicating that the star formation history of
fossil groups seems to be  analogous to that of field ellipticals.  We
further examined  elliptical galaxies in compact groups  studied by de
la Rosa  et al.   (2007).  As already  established in  previous works,
elliptical galaxies  in CGs are  older and more metal-poor  than field
ellipticals (Proctor et al.~2004; de  la Rosa et al.  2007) and fossil
groups.  Particularly, when we look  at the relation between [Z/H] and
log$\sigma$, the mass-metallicity relation  is evident (see Lee et al.
2006; Kobayashi 2005).   More massive galaxies (in both  FGs and field
galaxies) retain their metals more  efficiently than the low mass ones
(i.e., ellipticals in CGs).   The $\alpha$-enhancement diagram shows a
clear   trend   between   $\alpha/F_{e}$   and   velocity   dispersion
(mass). This can be interpreted as a manifestation of downsizing - the
less  massive galaxies have  a more  extended star  formation history.
Based  on  these  results  we  see  that  FGs  are  similar  to  field
ellipticals but  cannot be formed by  dry mergers of  ellipticals in a
CG. However, a wet merger with a gas rich disk system may also explain
the  observed  relations.   Here, the  luminosity-weighted  parameters
(Age,  [Z/H],   [$\alpha$/Fe])  are  more  sensitive   to  the  recent
star-formed-population, generating lower ages, higher [Z/H] and higher
[Mg/Fe].  This  assumes that the merger-starburst is  soon followed by
SN/AGN feedback which depletes the gas content.

McCarthy  et al.~(2004)  have found  a significant  population  of red
galaxies at high redshift ($1.3<z<2.2$), and claim that they cannot be
descendants of  the $z \sim  3$ Lyman-break galaxies.  Based  on their
spectroscopic data  they also  conclude that most  of the  present day
massive  galaxies   had  an  early   ($z_{\rm  f}=  2.4$)   and  rapid
formation. If we restrict ourselves to the twenty-five FGs, their last
main star formation  episode must have occurred at  $z>0.3$, much more
recently than in the high-$z$  galaxies. It appears that a more recent
burst  occurred in  the  FGs  which would  explain  their lower  ages.
Therefore, we may be seeing  massive ellipticals that accreted a small
galaxy at $z>0.3$.

Our  results indicate that  FGs are  not significantly  different from
bright ellipticals  found in low-density  regions of the  universe. As
claimed by  Mulchaey \&  Zabludoff (1999) in  a detailed study  of NGC
1132, FGs  could simply be "failed"  groups - a  few relatively bright
galaxies merged forming the dominant  system we see today, but without
enough surrounding  matter to form additional  nearby bright galaxies,
resulting in  an atypical LF.  This  is also in  agreement with recent
results    obtained    by   D'Onghia    et    al.~(2005)   based    on
N-body/hydrodynamical  cosmological simulations  where they  find that
FGs are formed by the infall of $L \sim L_{*}$ systems along filaments
with impact parameters as small as 5 kpc.

Our  findings  suggest  that  the objects  meeting  the  observational
criteria expected for fossil groups  at low redshift might not be true
fossils.   The extensive  study presented  here whereby  we  select FG
candidates  based  on optical  data,  and  then  examining their  X-ray
counterparts   reveals   no   difference   between   FGs   and   field
ellipticals. We  note that  fossils can be  created anytime  in cosmic
history,  and these  systems may  represent  the final  stage of  mass
assembly as suggested in the  analysis of the Millennium simulation by
Dariush et al. (2007), instead of forming a distinct class.





\appendix

\section{List of properties of FGs and field galaxies}
\label{tables}
Tab.~2 lists all the quantities used to characterize the samples of
FGs and field galaxies.  FGs with extended X-ray contours ($N=25$)
have flag=1 in column~21, while field galaxies ($N=17$) have flag=2.
The table also lists the four FG candidates that are not in
the final FG sample because they do not fulfill the X-ray extension
criterion.  These objects have flag=0.  Column 1 is a running number
for the list of FGs and field galaxies. Columns~2,~3, and~4 give the
RA and DEC (in degrees) and spectroscopic redshift from the SDSS DR4.
Columns~5,~6, and~7 provide the $z=0.1$ k-corrected absolute
magnitude, the magnitude gap of the FG, and the $g-r$ color,
respectively.  For field galaxies, the values of $\Delta M$ have been
randomly assigned as described in Sec.~\ref{sec:MAG_DIST}.  Column~8
provides the density excess around each galaxy.  Columns~9 and~10
report the effective radii (in arcsec) in the $g$ and $r$ bands, while
columns~11, and~12 list the corresponding Sersic indices.  Columns~13
and~14 provide the $a_4(\times 100)$ parameters and internal color
gradients (in $\rm mag/dex$).  Column~15 reports the X-ray luminosity
in units of $10^{44} h_{75}^{-2} \rm erg/s$. Column~16 provides the
distance of each galaxy to the red sequence (Eq.~\ref{eq:DIST_RS}).
Columns ~17 to~20 provide the velocity dispersions (in units of $\rm km/s$)
and stellar population parameters.

\section{Measuring the Red Sequence}
\label{subsec:RS_MEAS}

\begin{figure}[hb]
\begin{center}
\epsscale{0.5}
\plotone{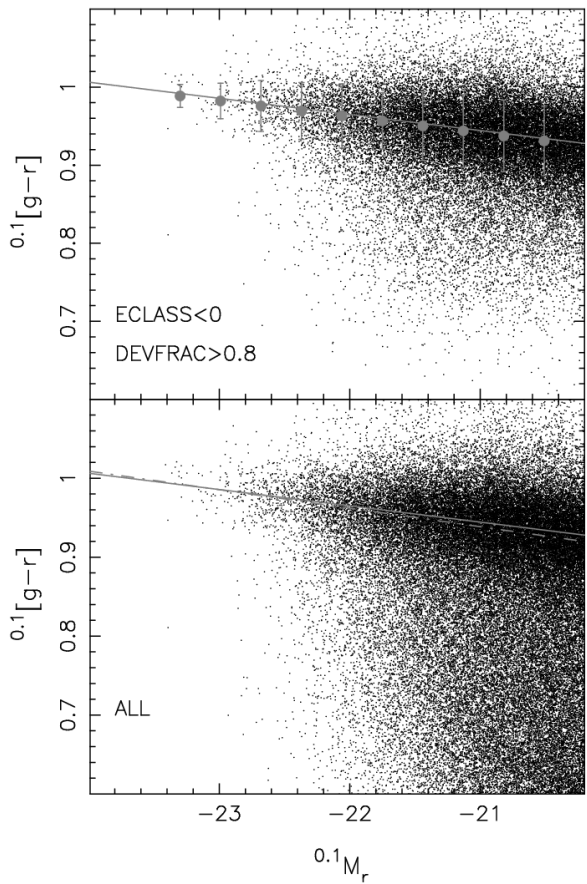}
\caption{Color--magnitude diagram of early-type galaxies (upper panel)
and of  all galaxies  (lower panel) in  the catalog. The
grey line  in the  upper panel  shows the bi-weight  fit to  the data,
while  grey  circles,  and  error  bars  plot  the  mean  and  standard
deviation,  $\sigma_{g-r}$,  of the  color  distribution in  different
magnitude bins. The grey solid line  in the lower panel is the same as
that  in  the  upper panel,  while  the  grey  dotted line  shows  the
bi-weight fit to the color--magnitude  relation of all galaxies in the
catalog.}
\label{CM}
\end{center}
\end{figure}

To quantify the  properties of the FG first-rank  galaxies relative to
the red sequence, we use  the SDSS-DR4 galaxy catalog to calculate the
offset, slope and scatter of the  $g-r$ vs. $r$ color-magnitude relation  
for all early-type galaxies.  Following a procedure similar to ~\citet{BER03a}, we select
early-type galaxies  on the basis of the  SDSS spectroscopic parameter
$eclass$,  which  classifies  the  spectral  type  using  a  principal
component analysis,  and the photometric  parameter $fracDev_r$, which
measures the fraction of the  galaxy light distribution that is better
fit by a  de Vaucouleurs law.  We define  as early-type galaxies those
systems  with $eclass  \!  < \!  0$  and $fracDev_r  \!   > \!   0.8$,
resulting in a sample of 39733 (out of 91563) galaxies selected from the SDSS catalog used to search for fossil systems (see Sec.~\ref{subsec:SEL}).  Total
galaxy  magnitudes are  taken from  the $r$-band  Petrosian magnitudes
output   by   the   SDSS   photometric  pipeline,   since   they   are
model-independent and highly reliable  for bright galaxies. Colors are
obtained from the $g$ and  $r$ band SDSS model magnitudes because they
are measured  in the same  aperture for all filters.   Both magnitudes
and  colors are k-corrected  using the  $ kcorrect$v4\_1\_4 software
package~\citep{BL03}, allowing  k-corrections to be  estimated through
filters that  are blue-shifted by a factor  $(1+z_0)$.  This procedure
has   the   primary   advantage   of   minimizing   uncertainties   on
k-corrections,  provided that  the  value  of $z_0$  is  close to  the
redshift range of the  observed galaxies.  Following other papers that
analyze the color  and magnitude distribution of galaxies  in the SDSS
database  (e.g.~\citealt{Hogg04}), we choose  $z_0=0.1$, corresponding
to the upper redshift limit of our galaxy catalog, and we indicate the
$z_0=0.1$ k-corrected  model magnitudes with  $^{0.1}g$ and $^{0.1}r$,
and the $z_0=0.1$  k-corrected Petrosian absolute magnitude  with the
notation $^{0.1}M_r$.

\begin{figure}
\begin{center}
\epsscale{0.4}
\plotone{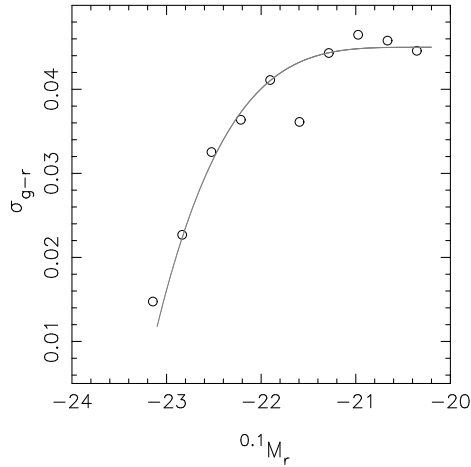}
\caption{The scatter around  the red sequence, $\sigma_{g-r}$, plotted
  as a  function of central value  of each magnitude bin  in the upper
  panel of Fig.~\ref{CM}. The grey  curve is the polynomial fit to the
  points, as given in Eq.~\ref{SGR}.}
\label{CMS}
\end{center}
\end{figure}

Fig.~\ref{CM}  shows the color--magnitude  diagram of  both early-type
and  all galaxies  in  our  catalog. Although  the  catalog is  almost
complete down  to $M_r=-20$ (see Sec.~\ref{subsec:SEL}),  we note that
the  difference   in  k-corrections   for  galaxies  with   $M_r  \sim
-20$\footnote{As  mentioned  in Sec.~\ref{sec:DENS_EX},  k-corrections
  have not been applied when  computing total magnitudes to select the
  SDSS-DR4 catalog.   } makes the  color--magnitude diagram incomplete
down to  that magnitude limit.   Therefore, we derive the  slope, $a$,
and the  offset, $b$, of  the color--magnitude relation  by minimizing
the  rms of  galaxy  colors  around the  linear  relation, using  only
galaxies brighter than $M_r \sim  -20.2$.  To reduce the effect
of   outliers,    the   color   rms   is    computed   via   bi-weight
statistics~\citep{Beers:90}.   The uncertainties  on $a$  and  $b$ are
estimated  through  the  bootstrap  method,  with  $N=1000$  bootstrap
iterations.  For  the sample of early-type galaxies,  we find $a=0.944
\pm  0.005$  and $b=-0.0206  \pm  0.003$,  consistent with  $b=-0.022$
reported by~\citet{HOGG03}  and with  other results in  the literature
(e.g.~\citealt{COOL06,  GCB06}).  The  derived values  of $a$  and $b$
change by less than $10 \%$  if we apply the same fitting procedure to
the  entire galaxy  catalog (i.e.  not  applying   the  $eclass$  and
$fracDev_r$  cuts),  demonstrating the  robustness  of the  regression
method.  To estimate the scatter around the color--magnitude relation,
we bin  the sample of  early-type galaxies with respect  to magnitude,
and derive the  color rms, $\sigma_{g-r}$, in each  bin with bi-weight
statistics.   Fig.~\ref{CMS}  plots $\sigma_{g-r}$  as  a function  of
$^{0.1}M_r$.  We  find that  $\sigma_{g-r}$ increases at  the faintest
magnitudes, with this trend well-fit by the relation
\begin{equation}
\sigma_{g-r}(^{0.1}M_r) = 0.045-0.00047 \times (^{0.1}M_r+20.2)^4,
\label{SGR}
\end{equation}
in  the  magnitude range $-23 < ^{0.1}M_r < -20.2$.   The  mean value  of
$\sigma_{g-r}$ is  $0.035$, consistent with that found in
previous studies (see e.g.~\citealt{COOL06}).

\section{Measuring structural parameters and internal color gradients}
\label{subsec:STRUCPAR}
\begin{figure}
\begin{center}
\epsscale{0.8}
\plotone{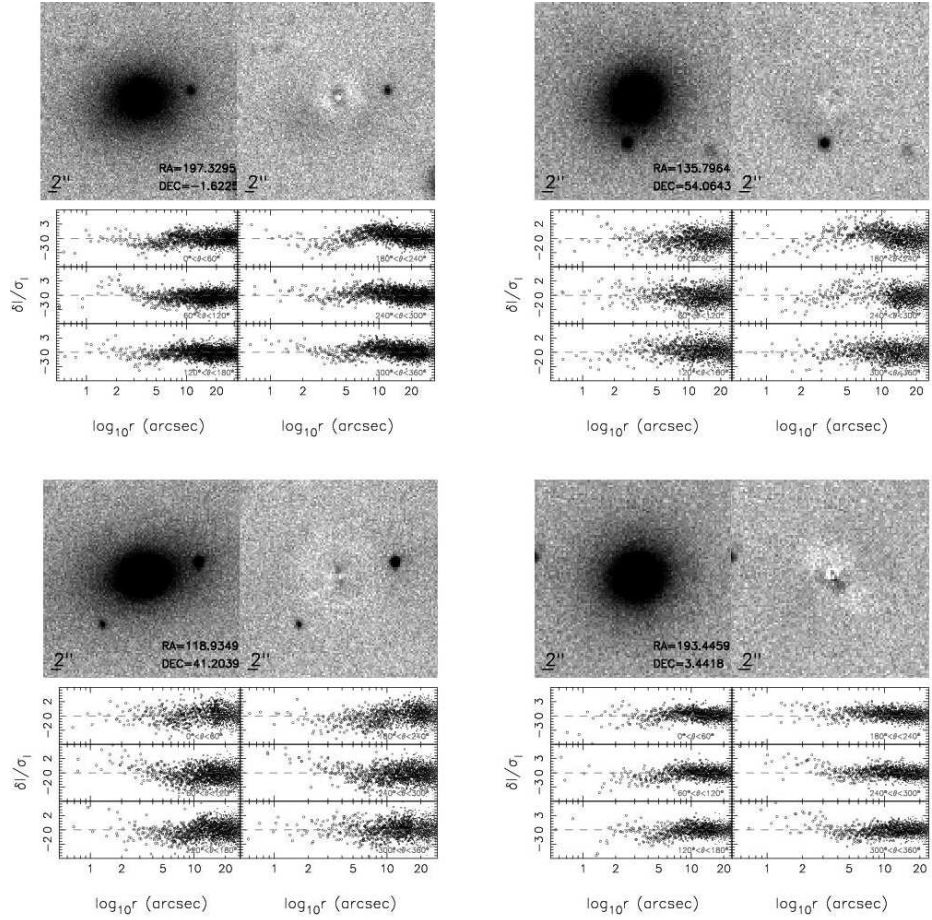}
\caption{  Two-dimensional  fitting of  four  galaxies  drawn from  the
  sample  of  twenty-five  FGs. For  each  galaxy,  the  upper--left and  right
  subpanels  show  the galaxy  $r$-band  image  and the  corresponding
  residual map after model subtraction, respectively.  Right ascension
  and declination are reported, in  degrees, in the lower right corner
  of the left subpanel. The  spatial scale is shown in the lower-left
  corners.  The  lower  six  subpanels  show  the  fitting  residuals,
  normalized to the  photon noise in each pixel, as  a function of the
  radial distance,  $r$, to the galaxy  center within six  bins of the
  polar  angle, $\theta$,  respectively. The  limits of  each  bin are
  shown in the lower right corner of each subpanel.  }
\label{fig:2DFIT}
\end{center}
\end{figure}

Structural  parameters  were  derived  using  2DPHOT  (La  Barbera  et
al.~2008,  hereafter LdC08),  an automated  software  environment that
performs several tasks, such as catalog extraction (using S-Extractor,
Bertin   \&   Arnouts~1996),   star/galaxy  separation,   and   surface
photometry.   To estimate  color  gradients,
structural parameters  were measured  in both the  $g$ and  $r$ bands.
For  each galaxy  in the  FG  and field (FS) samples  we  retrieved  the   
corresponding  $g$-  and  $r$-band  {\it
  corrected frames}  from the SDSS DR4 Data Archive  Server and
ran 2DPHOT on them.  A  complete description of the 2DPHOT package can
be found  in LdC08; here we  provide a brief description  of the steps
relevant to measuring galaxy  structural parameters.  For each galaxy,
a PSF model  was constructed by fitting the four  closest stars in the
image with a sum of three two-dimensional Moffat functions.  Isophotal
distortions of star isophotes were modeled as described in LdC08.  The
parameters $r_{\rm  e}$, $< \!\mu\!>_{\rm e  }$ and $\rm  n$ were then
obtained by  fitting galaxy  images with PSF-convolved  Sersic models.
As an example of the fitting procedure, Fig.~\ref{fig:2DFIT} plots the
images of a subset of galaxies from the list of twenty-five FGs, along with the
residuals after subtracting the best-fitting Sersic models.  The
shapes of FG and  FS candidates are characterized by
the  isophotal  shape parameter,  $a_4$,  of  each galaxy.   Isophotal
contours  were expanded into  a sin/cos  series~\citep{Bender:87}, and
analyzed as described  in LdC08, by deriving the  $a_4$ coefficient as
the average of the $a_4$ profile within a radial range from four times
the FWHM to twice the galaxy effective radius.

Internal  color gradients,  defined as  the logarithmic  slope  of the
galaxy  radial   color  profile,   were  estimated  as   follows  (see
also~\citealt{LaBarbera:03}).   For  each  galaxy  in the  FG  and  FS
samples,  the corresponding  $g$-  and $r$-  band best-fitting  Sersic
parameters were used to construct seeing-deconvolved galaxy images.  A
set  of concentric  ellipses were  constructed for  both the  $g$- and
$r$-band  galaxy images, with  all ellipses  having the  same position
angle and  axis ratio  parameters as those  derived from  the $r$-band
Sersic  fitting  (Sec.~\ref{subsec:STRUCPAR}).   For each  ellipse,  a
$g-r$  color was  calculated  as the  average  ratio of  the $g$-  and
$r$-band intensity  values along that ellipse. The  color profile was
then  computed  as  the average  $g-r$  color  as  a function  of  the
equivalent   radius  of  the   corresponding  ellipse.    An  ordinary
least-squares fit  to the color  profile, adopting the color  index as
the independent  variable, was then  performed, with the slope  of the
fitted  line yielding  the internal  color gradient  for  each galaxy.
Following previous  studies of color gradients  in early-type galaxies
(e.g.~\citealt{Peletier:90}), the fit was  performed in a range of the
color profile,  from an inner  radius of $0.1$  to an outer  radius of
$1.0$ times the $r$-band effective radius.

\section{Spectral analysis}
\label{app:spectra}
\subsection{Measuring spectral parameters}
The spectra  of FGs and field galaxies were masked to  avoid bad pixels affecting  the wavelength
ranges  defining the spectral  indices, brought  to redshift  zero and
corrected   for  foreground   dust   extinction  following   Schlegel,
Finkheiner  \&  Davis~1998.  Then,  we  derived  the spectral  indices
H$\beta$,  [MgFe]$^{\prime}$  =  (Mgb  (0.72 $\times$  Fe5270  +  0.28
$\times$ Fe5335))$^{1/2}$,  Fe3 = (Fe4383+Fe5270+Fe5335)/3,  and <Fe>=
(Fe5270+Fe5335)/2,  which are  defined  according to  the Lick  system
(Worthey \&  Ottaviani 1997; Gonz\'alez  1993), except for  the higher
resolution of the SDSS spectra.

\begin{figure}
\plotone{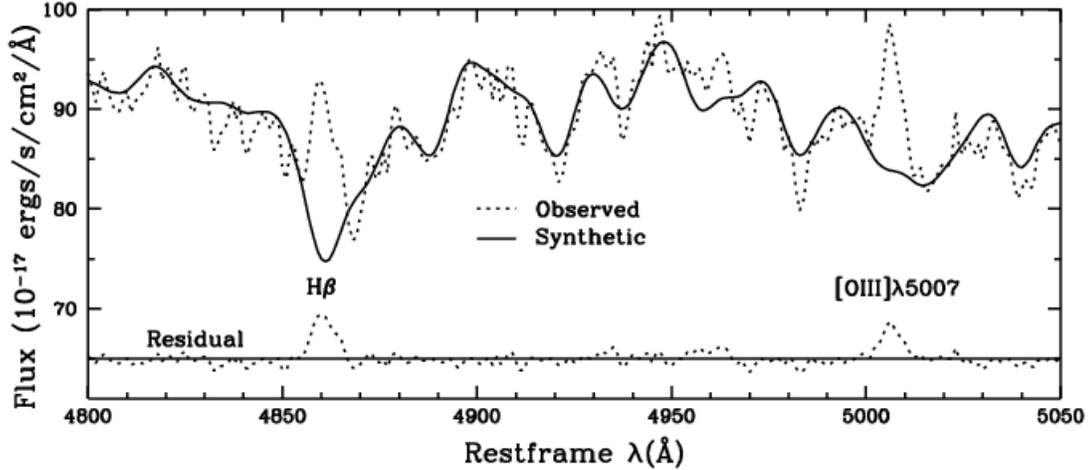}
\caption{The  synthetic  spectrum  obtained by  STARLIGHT  (continuous
  line) is the combination of  SSP models which best fits the observed
  SDSS-spectrum  (dotted  line). The  residual  spectrum (observed  --
  synthetic),  can be  used to  measure emission  indices.  Instead of
  correcting  for  emission filling,  we  measure  the H$\beta$  index
  directly from the synthetic spectrum}
\end{figure}

The  measurement of  the H$\beta$  spectral index,  which  is directly
related  to age, may  be significantly  affected by  nebular emission,
biasing the  inferred age towards  older values.  To correct  for this
effect, we used the STARLIGHT  code (Cid Fernandes et al.  2005) which
finds the  best fit  to the  observed spectrum by  combining a  set of
SSP-MILES   Spectral  Energy   Distributions  (SEDs;   see  Sec.~8.2),
convolved to the  SDSS spectral resolution and broadened  to match the
galaxy's velocity  dispersion.  As a result of  the fitting procedure,
STARLIGHT  also provides  the velocity  dispersion, $\sigma$,  of each
galaxy. During the fitting process, regions around emission lines were
masked.  Fig.~18 shows how the residual spectrum unveils the nebular
emission.  Instead  of correcting for  the residual emission  line, we
measured  the  H$\beta$ index  directly  from  the absorption  spectra
obtained by STARLIGHT.

Velocity dispersions of  FGs and field galaxies were  taken from the SDSS DR4.
For 53 galaxies (out of 102 FG candidates and 66 field galaxies with significant X-ray detection), the $\sigma$ measurements were
not available from the SDSS database,  and in this case we adopted the
STARLIGHT   estimates.   The   spectral  parameters,   i.e.,  velocity
dispersions and spectral indices,  were corrected to a common physical
aperture of  R$_{eff}$/8, following J\o rgensen  et al.~(1995) and~J\o
rgensen~(1997). For each galaxy,  the effective radius, R$_{eff}$, was
defined  as  $R_{eff} =  (b/a)\times  R_{dev}$,  using the  $deVRad_r$
($R_{dev}$) and  $deVAB_r$ ($b/a$) parameters from  the SDSS database.
Since  the aperture  corrections rely  on effective  radii  derived by
fitting  de  Vaucouleurs  models,  we  adopted  the  SDSS  photometric
pipeline's  $R_{eff}$ values,  instead  of using  our effective  radii
derived using Sersic models (see App.~\ref{subsec:STRUCPAR}).

\subsection{Extraction of stellar population parameters}
We measured  age, metallicity and abundance  ratio [$\alpha$/Fe] using
the recently developed $\alpha$-enhanced  SSP-MILES models (de la Rosa
et al.~2008),  which provide  SEDs at 2.4  \AA\ (FWHM)  resolution for
different  $\alpha$-enhancement ratios, combining  theoretical (Coelho
et al.~2005) and  empirical (MILES) libraries (S\'anchez-Bl\'azquez et
al.~2006).  For  each galaxy,  we constructed a  grid of  H$\beta$ and
[MgFe]$^{\prime}$  values  using  the  solar  metallicity  SSP  models
([$\alpha$/Fe]=0) with  a variety of age and  [Z/H] combinations.  The
age and [Z/H] of the  observed spectrum were obtained by interpolating
its two spectral indices among  those of the models.  The H$\beta$ and
[MgFe]$^{\prime}$ indices are mostly sensitive to age and metallicity,
respectively,  breaking  the  well-known age--metallicity  degeneracy.
Moreover,  they have  the  advantage of  being  almost insensitive  to
[$\alpha$/Fe]  (Korn, Maraston  \&  Thomas~2005).  As  a second  step,
after fixing the estimated age, we construct a new grid of Fe3 and Mgb
values  with  models  having  a  variety of  [Z/H]  and  [$\alpha$/Fe]
combinations.  Interpolation of the  observed indices yields the final
metallicity and [$\alpha$/Fe] values.  To check the reliability of the
$\alpha$-enhancement  estimates, we  repeated  the above  computation,
replacing the Fe3  index with $\rm < \!   Fe \!  >=(Fe5270+Fe5335)/2$.
The two  procedures are fully consistent  at a confidence  level of $2
\sigma$ for all but three of  the FG and FS galaxies.  For those three
objects, we adopted  the more accurate $\rm < \!  Fe  \!  > $ estimate
(see de la Rosa et al. ~2008).

\begin{acknowledgements}
We thank the referee for several comments that helped to significantly
improve  this paper.  We  acknowledge Dr.   D.Kocevski for  helping us
with X-ray analysis of the RASS data. G.S.  and A.R.  thank the EC and
the  MIUR  for having  partially  supported  this  work (EC  contracts
HPRN-CT-2002-00316    -    SISCO    network,    and    MIUR-COFIN-2004
n.2004020323\_001).\\  Funding  for  the  SDSS and  SDSS-II  has  been
provided  by  the  Alfred  P.   Sloan  Foundation,  the  Participating
Institutions, the National Science Foundation, the U.S.  Department of
Energy,  the  National   Aeronautics  and  Space  Administration,  the
Japanese  Monbukagakusho,  the  Max  Planck Society,  and  the  Higher
Education Funding  Council for  England.  The SDSS  is managed  by the
Astrophysical Research Consortium  for the Participating Institutions.
The  Participating Institutions  are  the American  Museum of  Natural
History,  Astrophysical   Institute  Potsdam,  University   of  Basel,
Cambridge University,  Case Western Reserve  University, University of
Chicago,  Drexel  University,  Fermilab,  the Institute  for  Advanced
Study, the  Japan Participation  Group, Johns Hopkins  University, the
Joint  Institute for  Nuclear  Astrophysics, the  Kavli Institute  for
Particle Astrophysics  and Cosmology, the Korean  Scientist Group, the
Chinese Academy of Sciences  (LAMOST), Los Alamos National Laboratory,
the Max-Planck-Institute for Astronomy (MPA), the Max-Planck-Institute
for  Astrophysics  (MPIA), New  Mexico  State  University, Ohio  State
University,  University  of   Pittsburgh,  University  of  Portsmouth,
Princeton  University, the  United States  Naval Observatory,  and the
University of Washington.
\end{acknowledgements}


\clearpage
\textheight=68\baselineskip

\begin{landscape}
\hspace{-1truecm}
\begin{deluxetable}{c|cccccccccccccccccccc}
\tablecaption{Properties of FG and FS galaxies. \hspace{-4cm}}
\tabletypesize{\tiny}
\tablewidth{0pt}
\tablecolumns{21}
\tablehead{ ID & R.A. & Dec & z & $^{0.1}M_r$ & $\Delta M$ & $^{0.1}g-r$ & $\delta_N$ & $r_{e,g}$ & $r_{e,r}$ & $n_{g}$ & $n_{r}$ & $a_4 (\times 100)$ & $\nabla(g-r)$ & $\log L_{\rm X}$ & $\delta_{\rm CM}$ & $\log \sigma_0 $  & $\log Age $ & $[Z/H]$ & $[\alpha/Fe]$ & FLAG \\ (1) & (2) & (3) & (4) & (5) & (6) & (7) & (8) & (9) & (10) & (11) & (12) & (13) & (14) & (15) & (16) & (17) & (18) & (19) & (20) & (21) } 
\startdata
1 & $ 197.32954$ & $   -1.6225$ & $  0.083 $ & $-22.917 $ & $   2.62 $ & $  1.020 $ & $   2.6 $ & $  9.11 $ & $  8.09 $ & $   4.1 $ & $   3.0 $ & $   0.24 $ & $ -0.068 $ & $   0.09 $ & $  1.894$ & $  2.494 $ & $    0.7 $ & $   0.47 $ & $   0.30 $ & $1 $ \\ 
2 &  $ 135.79644$ & $   54.0643$ & $  0.083 $ & $-22.663 $ & $   2.04 $ & $  0.988 $ & $   3.0 $ & $  4.14 $ & $  5.36 $ & $   4.0 $ & $   3.8 $ & $  -0.20 $ & $ -0.061 $ & $  -1.00 $ & $  0.360$ & $  2.479 $ & $    0.7 $ & $   0.50 $ & $   0.28 $ & $1 $ \\
3 &  $ 118.93488$ & $   41.2039$ & $  0.074 $ & $-22.908 $ & $   2.35 $ & $  0.965 $ & $   3.3 $ & $  5.66 $ & $  5.82 $ & $   5.4 $ & $   4.5 $ & $  -0.26 $ & $ -0.024 $ & $  -1.05 $ & $ -0.935$ & $  2.404 $ & $    0.9 $ & $   0.26 $ & $   0.20 $ & $1 $ \\
4 &  $ 193.44587$ & $    3.4418$ & $  0.066 $ & $-22.349 $ & $   2.41 $ & $  0.802 $ & $  -0.8 $ & $  4.52 $ & $  4.18 $ & $   5.4 $ & $   4.3 $ & $   0.14 $ & $  0.042 $ & $  -0.60 $ & $ -4.846$ & $  2.504 $ & $    0.4 $ & $   0.44 $ & $   0.24 $ & $1 $ \\
5 &  $ 242.70645$ & $   48.3879$ & $  0.090 $ & $-22.203 $ & $   1.91 $ & $  0.956 $ & $  -0.8 $ & $ 45.14 $ & $102.19 $ & $   8.0 $ & $   9.8 $ & $  -0.30 $ & $  0.093 $ & $  -0.27 $ & $ -0.338$ & $  2.317 $ & $    0.6 $ & $   0.42 $ & $   0.19 $ & $1 $ \\
6 &  $ 202.96613$ & $   -2.8721$ & $  0.086 $ & $-22.556 $ & $   2.34 $ & $  0.961 $ & $   3.9 $ & $  4.49 $ & $  6.11 $ & $   5.8 $ & $   5.5 $ & $   0.99 $ & $ -0.041 $ & $  -0.76 $ & $ -0.490$ & $  2.377 $ & $    0.7 $ & $   0.41 $ & $   0.26 $ & $1 $ \\
7 &  $ 171.65402$ & $   55.3564$ & $  0.070 $ & $-22.528 $ & $   1.90 $ & $  0.960 $ & $   1.5 $ & $300.11 $ & $ 22.26 $ & $   6.3 $ & $   4.2 $ & $  -0.28 $ & $ -1.155 $ & $  -0.90 $ & $ -0.509$ & $  2.414 $ & $    0.8 $ & $   0.55 $ & $   0.31 $ & $1 $ \\
8 &  $ 215.39972$ & $   44.7081$ & $  0.091 $ & $-22.781 $ & $   2.44 $ & $  1.006 $ & $   2.8 $ & $ 26.16 $ & $ 49.60 $ & $   8.6 $ & $   8.6 $ & $   0.97 $ & $  0.089 $ & $  -0.80 $ & $  1.041$ & $  2.483 $ & $    0.5 $ & $   0.57 $ & $   0.19 $ & $1 $ \\
9 &  $ 184.34689$ & $    8.8729$ & $  0.094 $ & $-22.785 $ & $   2.05 $ & $  0.952 $ & $   3.0 $ & $ 37.91 $ & $ 27.49 $ & $   3.9 $ & $   3.2 $ & $  -0.49 $ & $ -0.487 $ & $  -0.53 $ & $ -1.194$ & $  2.427 $ & $    0.7 $ & $   0.24 $ & $   0.47 $ & $1 $ \\
10 & $ 133.05354$ & $   29.4108$ & $  0.086 $ & $-22.959 $ & $   2.63 $ & $  0.970 $ & $   2.3 $ & $ 13.98 $ & $ 35.33 $ & $   2.3 $ & $   3.4 $ & $   1.29 $ & $  0.297 $ & $  -0.76 $ & $ -0.781$ & $  2.399 $ & $    0.7 $ & $   0.38 $ & $   0.20 $ & $1 $ \\
11 & $ 216.32744$ & $   40.4898$ & $  0.083 $ & $-22.035 $ & $   1.97 $ & $  0.964 $ & $  -0.0 $ & $ 10.82 $ & $ 12.82 $ & $   4.8 $ & $   4.5 $ & $   0.48 $ & $ -0.115 $ & $  -1.70 $ & $ -0.038$ & $  2.326 $ & $    0.7 $ & $   0.47 $ & $   0.23 $ & $1 $ \\
12 & $ 191.71340$ & $    0.2970$ & $  0.089 $ & $-23.248 $ & $   2.94 $ & $  1.006 $ & $   4.0 $ & $ 20.44 $ & $ 14.73 $ & $   7.5 $ & $   6.0 $ & $  -0.56 $ & $ -0.271 $ & $  -0.65 $ & $  1.150$ & $  2.505 $ & $    0.6 $ & $   0.36 $ & $   0.17 $ & $1 $ \\
13 & $ 134.52059$ & $    1.1537$ & $  0.071 $ & $-22.262 $ & $   1.94 $ & $  0.997 $ & $   2.2 $ & $ 26.81 $ & $ 25.39 $ & $   9.4 $ & $   9.0 $ & $  -0.03 $ & $ -0.108 $ & $  -1.19 $ & $  0.727$ & $  2.404 $ & $    0.8 $ & $   0.64 $ & $   0.35 $ & $1 $ \\
14 & $ 117.99395$ & $   20.7490$ & $  0.077 $ & $-21.775 $ & $   1.83 $ & $  0.954 $ & $   1.3 $ & $  7.37 $ & $  8.38 $ & $   4.5 $ & $   4.2 $ & $   1.14 $ & $ -0.102 $ & $  -1.04 $ & $ -0.145$ & $  2.282 $ & $    0.5 $ & $   0.53 $ & $   0.16 $ & $1 $ \\
15 & $   9.36782$ & $   15.6263$ & $  0.080 $ & $-22.800 $ & $   2.42 $ & $  0.987 $ & $   2.0 $ & $ 11.72 $ & $ 10.64 $ & $   6.4 $ & $   5.1 $ & $  -0.06 $ & $ -0.096 $ & $  -1.15 $ & $  0.260$ & $  2.371 $ & $    0.9 $ & $   0.10 $ & $   0.11 $ & $1 $ \\
16 & $ 133.05970$ & $   29.3385$ & $  0.085 $ & $-22.196 $ & $   1.89 $ & $  0.993 $ & $   2.3 $ & $  8.64 $ & $ 12.85 $ & $   6.1 $ & $   6.5 $ & $   0.15 $ & $ -0.046 $ & $  -1.04 $ & $  0.646$ & $  2.457 $ & $    0.8 $ & $   0.74 $ & $   0.25 $ & $1 $ \\
17 & $ 325.61206$ & $   -6.5868$ & $  0.088 $ & $-22.750 $ & $   2.28 $ & $  0.979 $ & $   3.4 $ & $ 26.36 $ & $ 35.56 $ & $   9.9 $ & $  10.2 $ & $   2.42 $ & $ -0.070 $ & $  -0.93 $ & $ -0.050$ & $  2.409 $ & $    0.9 $ & $   0.19 $ & $   0.09 $ & $1 $ \\
18 & $ 358.77843$ & $   -9.3756$ & $  0.074 $ & $-22.516 $ & $   2.49 $ & $  0.981 $ & $   4.2 $ & $  4.28 $ & $  4.92 $ & $   3.6 $ & $   3.6 $ & $   0.66 $ & $ -0.139 $ & $  -0.98 $ & $  0.168$ & $  2.387 $ & $    0.7 $ & $   0.35 $ & $   0.41 $ & $1 $ \\ 
19 & $ 188.89591$ & $    1.8937$ & $  0.078 $ & $-22.520 $ & $   2.16 $ & $  1.025 $ & $   3.6 $ & $ 37.61 $ & $ 25.27 $ & $   8.6 $ & $   7.0 $ & $  -0.01 $ & $ -0.223 $ & $  -1.04 $ & $  1.568$ & $  2.520 $ & $    0.7 $ & $   0.47 $ & $   0.25 $ & $1 $ \\ 
20 & $ 195.28142$ & $   -3.4480$ & $  0.083 $ & $-22.761 $ & $   2.74 $ & $  1.007 $ & $   5.6 $ & $ 10.16 $ & $ 12.72 $ & $   4.9 $ & $   4.7 $ & $   0.46 $ & $ -0.086 $ & $  -1.11 $ & $  1.070$ & $  2.425 $ & $    0.9 $ & $   0.57 $ & $   0.34 $ & $1 $ \\ 
21 & $ 354.57353$ & $   15.6689$ & $  0.066 $ & $-22.600 $ & $   2.68 $ & $  0.973 $ & $   1.7 $ & $  8.45 $ & $ 14.56 $ & $   2.7 $ & $   4.0 $ & $   2.66 $ & $  0.044 $ & $  -1.77 $ & $ -0.122$ & $  2.361 $ & $    0.8 $ & $   0.21 $ & $   0.23 $ & $1 $ \\ 
22 & $ 129.81595$ & $   28.8441$ & $  0.079 $ & $-23.281 $ & $   3.04 $ & $  0.993 $ & $   4.6 $ & $  4.51 $ & $  6.14 $ & $   2.1 $ & $   2.4 $ & $   0.20 $ & $ -0.203 $ & $  -1.10 $ & $  0.122$ & $  2.543 $ & $    0.7 $ & $   0.78 $ & $   0.38 $ & $1 $ \\ 
23 & $ 230.21770$ & $   48.6607$ & $  0.074 $ & $-23.126 $ & $   3.15 $ & $  0.943 $ & $   4.0 $ & $ 10.66 $ & $ 12.67 $ & $   4.7 $ & $   4.8 $ & $  -0.72 $ & $ -0.102 $ & $  -0.40 $ & $ -3.176$ & $  2.529 $ & $    1.0 $ & $   0.27 $ & $   0.45 $ & $1 $ \\ 
24 & $ 228.00073$ & $   36.6520$ & $  0.066 $ & $-22.827 $ & $   2.91 $ & $  0.980 $ & $   0.9 $ & $ 11.04 $ & $ 14.23 $ & $   3.2 $ & $   3.3 $ & $  -1.16 $ & $  0.059 $ & $  -0.49 $ & $ -0.078$ & $  2.417 $ & $    0.7 $ & $   0.60 $ & $   0.14 $ & $1 $ \\ 
25 & $ 234.78742$ & $   33.6881$ & $  0.070 $ & $-22.555 $ & $   1.88 $ & $  1.016 $ & $   3.8 $ & $ 20.37 $ & $ 20.19 $ & $   9.9 $ & $  10.4 $ & $   0.09 $ & $ -0.129 $ & $  -0.84 $ & $  1.299$ & $  2.458 $ & $    0.7 $ & $   0.46 $ & $   0.45 $ & $1 $ \\ 
26 &$ 172.88388$ & $   12.6997$ & $  0.081 $ & $-22.306 $ & $   1.85 $ & $  0.972 $ & $   1.0 $ & $  7.00 $ & $  9.99 $ & $   2.4 $ & $   2.6 $ & $  -0.04 $ & $ -0.108 $ & $  -1.23 $ & $  0.017$ & $  2.434 $ & $    0.4 $ & $   1.46 $ & $   0.31 $ & $0 $ \\ 
27 & $  20.09644$ & $   -0.0790$ & $  0.077 $ & $-23.128 $ & $   3.01 $ & $  1.003 $ & $   1.6 $ & $ 32.20 $ & $ 39.10 $ & $   4.8 $ & $   5.0 $ & $  -0.42 $ & $ -0.133 $ & $  -1.00 $ & $  1.110$ & $  2.543 $ & $    0.8 $ & $   0.40 $ & $   0.32 $ & $0 $ \\
28 & $ 180.40768$ & $    1.1980$ & $  0.082 $ & $-22.225 $ & $   2.14 $ & $  0.903 $ & $  -2.6 $ & $ 10.41 $ & $  9.42 $ & $   4.7 $ & $   9.2 $ & $   2.29 $ & $ -0.848 $ & $  -0.85 $ & $ -1.773$ & $  2.197 $ & $    0.6 $ & $   0.09 $ & $   0.03 $ & $0 $ \\
29 & $ 118.18416$ & $   45.9493$ & $  0.052 $ & $-22.167 $ & $   2.05 $ & $  0.963 $ & $   2.3 $ & $  6.06 $ & $  5.78 $ & $   4.5 $ & $   4.8 $ & $  -0.45 $ & $ -0.042 $ & $  -0.96 $ & $ -0.144$ & $  2.336 $ & $    0.7 $ & $   0.23 $ & $   0.17 $ & $0 $ \\
30 &$ 168.84945$ & $   54.4441$ & $  0.070 $ & $-22.802 $ & $   1.99 $ & $  0.939 $ & $   2.6 $ & $  7.22 $ & $  7.38 $ & $   6.3 $ & $   5.6 $ & $  -0.89 $ & $ -0.041 $ & $  -1.42 $ & $ -1.795$ & $  2.349 $ & $    1.0 $ & $   0.09 $ & $   0.25 $ & $2 $ \\ 
31 & $ 232.31104$ & $   52.8640$ & $  0.073 $ & $-22.927 $ & $   2.17 $ & $  0.995 $ & $   3.2 $ & $ 21.23 $ & $ 24.43 $ & $   9.8 $ & $   8.6 $ & $  -0.34 $ & $  0.017 $ & $  -0.36 $ & $  0.602$ & $  2.454 $ & $    0.8 $ & $   0.60 $ & $   0.41 $ & $2 $ \\
32 & $ 122.59694$ & $   42.2739$ & $  0.064 $ & $-22.719 $ & $   2.40 $ & $  1.013 $ & $   3.7 $ & $  4.56 $ & $  5.02 $ & $   2.4 $ & $   2.9 $ & $   0.10 $ & $ -0.321 $ & $  -1.31 $ & $  1.298$ & $  2.571 $ & $    0.9 $ & $   0.63 $ & $   0.21 $ & $2 $ \\
33 & $ 184.62019$ & $   42.4610$ & $  0.073 $ & $-22.029 $ & $   1.93 $ & $  0.925 $ & $  -0.4 $ & $  9.19 $ & $  9.73 $ & $   5.1 $ & $   4.6 $ & $  -0.05 $ & $ -0.054 $ & $  -1.23 $ & $ -1.015$ & $  2.284 $ & $    0.9 $ & $   0.15 $ & $   0.17 $ & $2 $ \\
34 & $ 254.08789$ & $   39.2752$ & $  0.062 $ & $-22.595 $ & $   2.42 $ & $  0.976 $ & $   3.2 $ & $294.65 $ & $ 50.08 $ & $  10.7 $ & $  10.1 $ & $  -1.20 $ & $ -0.700 $ & $  -1.01 $ & $ -0.038$ & $  2.422 $ & $    0.6 $ & $   0.48 $ & $   0.31 $ & $2 $ \\
35 & $ 129.46447$ & $   26.5988$ & $  0.088 $ & $-22.705 $ & $   1.88 $ & $  0.903 $ & $   2.2 $ & $  6.32 $ & $  8.82 $ & $   6.2 $ & $   6.3 $ & $  -0.39 $ & $ -0.069 $ & $  -0.99 $ & $ -2.875$ & $  2.413 $ & $    0.5 $ & $   0.92 $ & $   0.09 $ & $2 $ \\
36 & $ 134.49123$ & $   30.3439$ & $  0.086 $ & $-22.817 $ & $   2.26 $ & $  0.949 $ & $   1.2 $ & $ 15.11 $ & $ 14.76 $ & $   7.4 $ & $   6.5 $ & $  -0.10 $ & $ -0.147 $ & $  -0.57 $ & $ -1.422$ & $  2.439 $ & $    0.9 $ & $   0.11 $ & $   0.41 $ & $2 $ \\
37 & $ 133.51888$ & $   29.0535$ & $  0.084 $ & $-22.938 $ & $   2.50 $ & $  1.002 $ & $   4.6 $ & $  8.06 $ & $ 13.29 $ & $   2.9 $ & $   3.6 $ & $   1.72 $ & $ -0.099 $ & $  -0.49 $ & $  0.964$ & $  2.463 $ & $    0.9 $ & $   0.35 $ & $   0.36 $ & $2 $ \\
38 & $ 237.97868$ & $   27.8640$ & $  0.082 $ & $-21.860 $ & $   1.81 $ & $  0.979 $ & $   2.8 $ & $  8.80 $ & $ 13.18 $ & $   2.9 $ & $   4.1 $ & $  -0.01 $ & $ -0.327 $ & $   0.29 $ & $  0.421$ & $  1.996 $ & $    0.7 $ & $   0.56 $ & $   0.29 $ & $2 $ \\
39 & $ 136.68638$ & $    3.6002$ & $  0.072 $ & $-22.123 $ & $   1.90 $ & $  0.893 $ & $  -1.5 $ & $ 12.94 $ & $ 21.63 $ & $  10.9 $ & $  10.2 $ & $   0.97 $ & $  0.146 $ & $  -0.51 $ & $ -1.928$ & $  2.209 $ & $    0.5 $ & $  -0.07 $ & $   0.05 $ & $2 $ \\
40 & $ 144.40096$ & $    7.9180$ & $  0.093 $ & $-21.778 $ & $   1.75 $ & $  0.925 $ & $  -1.3 $ & $  2.70 $ & $  4.25 $ & $   4.2 $ & $   4.4 $ & $   0.91 $ & $ -0.050 $ & $  -1.40 $ & $ -0.824$ & $  2.270 $ & $    0.5 $ & $   0.20 $ & $   0.45 $ & $2 $ \\
41 & $ 141.65320$ & $    3.4606$ & $  0.088 $ & $-22.235 $ & $   2.15 $ & $  0.954 $ & $   3.3 $ & $  4.45 $ & $  7.72 $ & $   4.8 $ & $   5.9 $ & $  -0.06 $ & $ -0.075 $ & $  -1.24 $ & $ -0.427$ & $  2.422 $ & $    0.9 $ & $   0.21 $ & $   0.19 $ & $2 $ \\
42 & $ 178.15655$ & $    3.4727$ & $  0.081 $ & $-22.771 $ & $   2.22 $ & $  0.964 $ & $   3.0 $ & $  5.71 $ & $  9.31 $ & $   1.6 $ & $   2.9 $ & $   0.41 $ & $ -0.625 $ & $   0.03 $ & $ -0.665$ & $  2.417 $ & $    0.8 $ & $   0.54 $ & $   0.50 $ & $2 $ \\
43 & $   7.36845$ & $   -0.2126$ & $  0.060 $ & $-22.322 $ & $   1.85 $ & $  0.981 $ & $   5.3 $ & $  4.82 $ & $  4.83 $ & $   2.7 $ & $   2.5 $ & $   0.47 $ & $  0.019 $ & $  -1.35 $ & $  0.288$ & $  2.390 $ & $    0.8 $ & $   0.41 $ & $   0.32 $ & $2 $ \\
44 & $ 216.19754$ & $    2.6644$ & $  0.053 $ & $-22.400 $ & $   1.95 $ & $  0.961 $ & $   3.4 $ & $  7.04 $ & $  6.69 $ & $   4.5 $ & $   4.5 $ & $   1.04 $ & $ -0.009 $ & $  -0.90 $ & $ -0.344$ & $  2.401 $ & $    0.8 $ & $   0.34 $ & $   0.25 $ & $2 $ \\
45 & $ 223.93717$ & $   -0.3062$ & $  0.083 $ & $-22.295 $ & $   1.92 $ & $  0.905 $ & $   1.7 $ & $  1.09 $ & $  1.49 $ & $   2.0 $ & $   2.4 $ & $   0.91 $ & $ -0.369 $ & $  -0.81 $ & $ -1.818$ & $  2.440 $ & $    0.8 $ & $   0.24 $ & $   0.20 $ & $2 $ \\
46 & $ 350.73485$ & $  -10.0456$ & $  0.084 $ & $-22.368 $ & $   2.24 $ & $  0.989 $ & $   5.6 $ & $  4.94 $ & $ 10.10 $ & $   5.6 $ & $   7.8 $ & $   1.13 $ & $  0.015 $ & $  -1.10 $ & $  0.477$ & $  2.444 $ & $    0.9 $ & $   0.32 $ & $   0.36 $ & $2 $ \\
\enddata
\end{deluxetable}
\clearpage
\end{landscape}
\textheight=64\baselineskip

\end{document}